\RequirePackage{fix-cm}

\documentclass[pdftex,onecolumn,epjc3]{preprint}   

\usepackage{graphicx}
\usepackage[intlimits,sumlimits,centertags]{amsmath}
\usepackage{amssymb,amsfonts}
\usepackage{gensymb}
\usepackage[bookmarksopen,bookmarksdepth=1]{hyperref}
\usepackage[x11names]{xcolor}
\usepackage{mathpazo} 
\usepackage{microtype}
\usepackage[a4paper, left=3.5cm,right=3.5cm,top=3.5cm,bottom=4cm]{geometry}
\usepackage[normalem]{ulem}
\usepackage{textcomp}
\usepackage{cite}

\journalname{Symmetry}

\begin{document}

\title{\huge Status, Challenges and Directions in Indirect Dark Matter Searches}

\author{\Large Carlos P\'erez de los Heros\thanksref{email03}}

\thankstext{email03}{e-mail: \href{mailto:cph@physics.uu.se}{cph@physics.uu.se}}

\institute{\large 
Department of Physics and Astronomy, Uppsala University, S-75120 Uppsala, Sweden}
\date{}

\maketitle

\begin{abstract}Indirect searches for dark matter are based on detecting an~anomalous flux of photons, neutrinos or cosmic-rays produced in annihilations or decays of dark matter candidates gravitationally accumulated in heavy cosmological objects, like galaxies, the Sun or the Earth. Additionally, evidence for dark matter that can also be understood as indirect can be obtained from early universe probes, like fluctuations of the cosmic microwave background temperature, the~primordial abundance of light elements or the Hydrogen 21-cm line.  
  The techniques needed to detect these different signatures require very different types of detectors: Air shower arrays, gamma-~and X-ray telescopes, neutrino telescopes, radio telescopes or particle detectors in balloons or satellites. While many of these detectors were not originally intended to search for dark matter, they have proven to be unique complementary tools for direct search efforts.  In this review we summarize the current status of indirect searches for dark matter, mentioning also the challenges and limitations that these techniques encounter.

\end{abstract}


\section{Introduction}
\label{sec:intro}

The term ``dark matter'' was originally coined because of the need to explain the observed rotation curves of stars in galaxies and the peculiar velocities
of galaxies within clusters. The~velocities of these large gravitational systems were measured to be incompatible with the expectations
based on Newtonian mechanics and the visible amount of matter in the system~\cite{Zwicky:1933gu,Bosma:1981zz,Sofue:2000jx}. These observations seemed to provide indirect evidence for the
existence of either a~non-luminous matter component in the universe, or~a deviation from the standard Newtonian mechanics as we use it to try
to make sense of the universe at larges scales~\cite{Castiblanco:2019mgb}. Additional matter is also needed to understand how 
the first galaxies could have formed from the small density perturbations imprinted in the cosmic microwave background (CMB)~\cite{Cyburt:2015mya} and to describe the large-scale
structure of the universe as seen today~\cite{Planelles:2014zaa,Vogelsberger:2019ynw}. Observations of gravitational lensing added to the list of evidence for the need of
dark matter~\cite{Ellis:2010aaa}.
While proposed modifications of gravity and other recent approaches~\cite{Famaey:2011kh,Deur:2017aas,Christodoulou:2015qna,Khoury:2014tka} do not need dark matter at all, in~this review we summarize the current experimental efforts
aimed at probing indirectly the most popular ``solution'' to the dark matter problem: The particle~solution.

\begin{figure}[t]
\centering\includegraphics[width=0.55\linewidth]{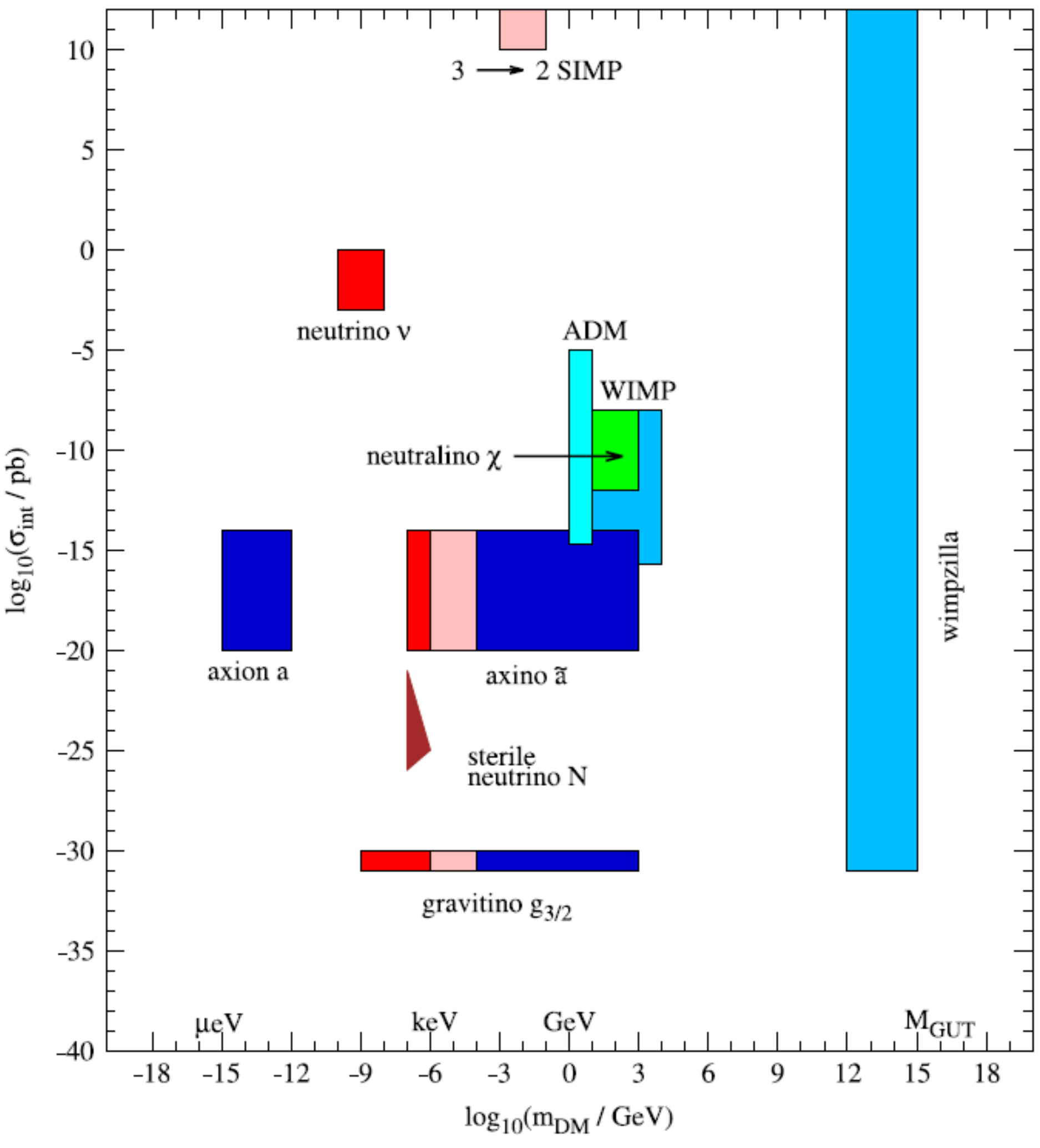} 
\caption{Popular candidates of dark matter indicating their mass and strength of interaction with ordinary matter. The~red, pink and blue colors represent
  hot dark matter (HDM), warm dark matter (WDM) and cold dark matter (CDM) candidates, respectively. Reprinted from~\cite{Baer:2014eja} with permission from Elsevier.}
\label{fig:candidates}
\end{figure}

 The particle solution to the dark matter problem is based on the assumption that stable (or enough long-lived) relic particles exist, whose present density is
 determined by the thermal history of the early universe~\cite{Steigman:2012nb}. Incidentally, models needed to explain certain shortcomings of the Standard Model of particle
 physics do provide viable candidates for  dark matter in the form of new weakly-interacting massive particles (WIMPs) in the GeV-$\mathcal{O}$(100~TeV)  mass range.
 If they are thermally produced in the early universe, the~upper mass limit for WIMPs arises from theoretical arguments in order to preserve unitarity~\cite{Blum:2014dca}.
 There is practically no lower limit on the mass of the dark matter candidates as long as they can perform their role of being abundant enough to solve
 the problem and they do not over close the universe~\cite{Leane:2018kjk}.  Higher masses than $\mathcal{O}$(100~TeV) and/or stronger interactions can be accommodated in models
 where the dark matter candidates are not produced thermally~\cite{Baer:2014eja}.  So the dark matter zoo encompasses a~wide class of particles with a~wide range of interactions
 as illustrated in Figure~\ref{fig:candidates}.  From~a theoretical point of view, the~connection between a~cosmological problem (the need for additional matter at cosmological
 scales) and the possibility of explaining it with particle physics ideas is extremely~compelling. 
 
The term ``indirect'' appertaining to dark matter searches refers to those methods based on detecting the products of the annihilation or
decays of dark matter candidates gravitationally accumulated in cosmological objects, in~contrast with ``direct'' detection techniques
where the interactions of dark matter particles from the local halo with a~suitable target are searched for~\cite{Schumann:2019eaa}.
Still, these two techniques rely on detecting the dark matter that surrounds us. Usually not included in ``indirect'' searches is the use of early universe
probes (cosmic microwave background power spectrum, 21-cm astronomy and nucleosynthesis) to set constraints to the dark matter content and characteristics. There is no
more indirect way of searching for something that using signatures that might have occurred billions of years ago, so we have included a~short description of dark matter
searches using those probes at the end of this review. 
 Additionally, accelerator searches try to produce new
particles in the controlled conditions of collider or fixed target interactions, that could be identified also as candidates for dark matter~\cite{Buchmueller:2017qhf,Kahlhoefer:2017dnp,Bondi:2020xnc}. In~any case, the~problem remains to identify and characterize this new form of matter, if~this is 
indeed the way Nature chose to~function.

Any review about dark matter, be it about direct or indirect searches, faces at least two, I think insurmountable, problems: To be sufficiently up to date in a~rapidly
evolving field, and~to give due credit to previous work by the innumerable scientists that have contributed to the field. I confess already here that I have given up in trying to solve these
two problems. Instead, I have tried to provide a~personal choice of topics and mention some general problems faced in the field of indirect dark matter searches 
that I think are relevant and that sometimes can have a~bearing on the interpretation of results (the choice of halo model or the consideration of particle physics or astrophysical uncertainties 
for example). This means that I have not followed a~historical approach (there are excellent reviews from that perspective~\cite{Sanders:2010aaa,Bertone:2016nfn}), neither I have tried to be complete in describing the
theoretical models being usually, or~less usually, considered (there are also excellent reviews on that~\cite{Profumo:2019ujg,Feng:2010gw,Bertone:2004pz}), or~tried to be comprehensive in mentioning the
results of every single experiment that exists in the field. I~have also favoured citing published results than preprints or conference proceedings when possible, sometimes~at the expense of not pointing the reader to the very latest results. With~these caveats in mind, I hope this review is of use to some~readers.

 \section{Dark matter in galactic halos}
 \label{sec:halo}

 A viable particle candidate for dark matter must be stable in timescales comparable with the age of the universe and have interactions with ordinary matter such it does
 not affect the evolution of the universe in a~way that leads to a~large-scale structure that is incompatible with observations today. It must also be neutral,
 otherwise star formation and evolution would have been disturbed in ways incompatible with observations~\cite{Gould:1989gw}, and~the formation of ``dark atoms''
 in the interstellar medium should have already been revealed in spectroscopy studies of heavenly objects~\cite{Basdevant:1989fh}. However,~micro-charges, about
 a~million times smaller than the charge on the electron, can still be accommodated if only a~few percent of the dark matter would have such charge~\cite{Munoz:2018pzp}. 
 Aside these generic constraints, there are still plenty of room for dark matter candidates with characteristics that avoid these limitations. And~there are plenty of
 theoretical models that can accommodate them.  Even~though WIMPs are popular candidates because straightforward extensions of the Standard Model,
 like many flavours of super-symmetry, predict them~\cite{Jungman:1995df}, extremely light dark matter in the
 form of axion-like particles (ALPs)~\cite{Pargner:2019wxt} or super-heavy dark matter~\cite{Chung:1998zb,Chung:1998ua} are also well justified. We will therefore not use the term WIMP in this
 review since many of the experimental results summarized later can be reinterpreted in terms of more general dark matter candidates. See~\cite{Palomares-Ruiz:2020ytu,Plehn:2017fdg,Khlopov:2018ttr,Zhang:2015ffa,Baer:2014eja} 
 and references therein for recent reviews of candidates and production scenarios beyond the vanilla WIMP~miracle.
 
 \begin{figure}[t]
  \centering\includegraphics[width=0.7\linewidth]{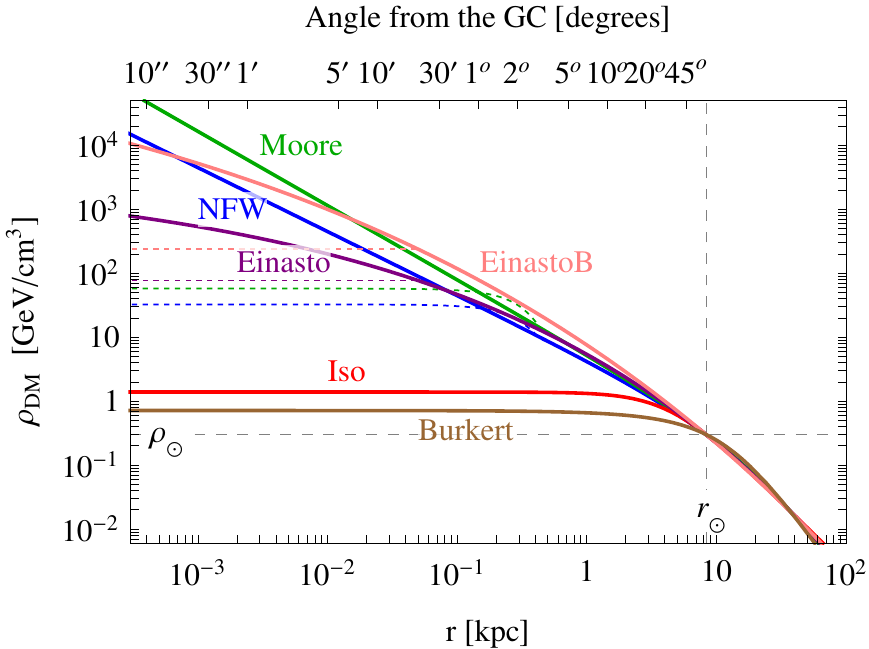} 
  \caption{Commonly used Milky Way dark matter density profiles, including an~isothermal model (Iso). See the
    reference below for details on the parameters used to obtain each curve. Here, suffice to show that the choice of one profile or another can have important consequences on the
    interpretation of experimental results. The~distributions are normalized to a~local density $\rho_{\odot}=0.3$ GeV/cm$^3$ in the solar neighborhood,
    at $r_{\odot}=8.3$ kpc from the Galactic Centre. Figure from~\cite{Cirelli:2010xx}, reproduced by permission of IOP Publishing. Copyright SISSA Medialab Srl. All rights reserved.
}
\label{fig:DMprofiles}
\end{figure}

Accumulations of dark matter are expected in halos around galaxies, including the Milky Way, and~in galaxy clusters. 
 In order to detect signatures from these regions with high density of dark matter, the~dark matter particles need to be able to annihilate between themselves, or~decay. 
 It is the stable standard model particles that result from these processes that constitute the potential signature for indirect searches. 
 These can be neutrinos, photons, electrons, protons or even light nuclei like He and D (and their antiparticles).
 
 The distribution of dark matter in the halos of galaxies is an~important ingredient both in calculating expected observables and eventually in 
 interpreting a~signal~\cite{Salucci:2018hqu,Zavala:2019gpq}. We know from N-body simulations of galaxy formation that galaxies are embedded in clumpy, rather complex dark matter halos
 that extend well beyond the visible galaxy~\cite{Springel:2005nw}. There have been several proposals to describe the dark matter density distribution in galactic halos as a~function of
 the distance from the centre of the galaxy~\cite{Gunn:1977aa,Navarro:1995iw,Einasto:1965czb,Salucci:2000ps,Moore:1999gc,Kravtsov:1997dp}. The~common feature of these profiles is a~denser,
 typically spherically symmetric, region of dark matter in the centre of the galaxy, with~decreasing density as the radial distance to the centre increases. 
Figure~\ref{fig:DMprofiles} shows a~few examples of these profiles. 
 Note that the slope of the density distribution varies with the distance, deviating from initial calculations in~\cite{Gunn:1977aa} that predicted a~single power law distribution
 across the whole distance range.  
 Where the different models diverge is in the predicted shape of the central region. Profiles obtained from N-body simulations of galaxy formation and 
evolution with a~collisionless dark matter contribution tend to predict a~steep power-law type behaviour of the dark matter component in the central region~\cite{Navarro:1995iw,Moore:1999gc}, while 
profiles based on observational data (stellar velocity fields) tend to favour a~more constant dark matter density near the core~\cite{Einasto:1965czb,Salucci:2000ps,Kravtsov:1997dp,Burkert:1995yz}.
This discrepancy has been dubbed the core-cusp problem~\cite{deBlok:2009sp}, and~it is an~unresolved issue. A~recent fit to the motion of stars in the disk of the Milky Way~\cite{Benito:2019ngh}
gives inconclusive evidence for a~cored or cusped profile, reflecting on the observational difficulty in accurate mapping the dark matter distribution even of our galaxy. There are proposals
within the dark matter paradigm that can ameliorate the problem, or~even solve it, based on self-interacting dark matter models~\cite{Spergel:1999mh,Rocha:2012jg,Peter:2012jh,Vogelsberger:2012ku,Bernal:2019uqr}. Self-interactions among dark matter particles
while dark matter halos form will heat up the dark matter component, avoiding the otherwise strong collapse that forms cusped halos. The~strength of the interactions (and therefore the mean free path of the
dark matter particles between interactions) can be chosen to reproduce observations, including avoiding undesirable effects in early universe cosmology. Particle physics models describing the
type of interactions that are allowed can be built under these simple assumptions. 
However, this solution still suffers from the fact that no dark matter candidates have been identified. A~less exotic solution to the cusp problem could arise from the fact that one can actually
describe the inner dark core from kinematic measurements of the visible inner region and global characteristics of the galaxy, like its viral mass~\cite{Salucci:2007tm}. This provides a
measurement of the size of the dark matter core which is directly correlated to the luminous counterpart and that does not show any cuspy behaviour. A~more radical claim is the recent study
in~\cite{Baushev:2018ptw} proposing that the cusp
obtained in N-body simulations represents a~numerical issue of the simulations themselves that do not reflect reality. Lacking further confirmation of the claims of the authors, their results
can be taken to illustrate the difficulties and limitations involved in performing N-body simulations, even~with today's powerful~computers.

Following~\cite{Hernquist:1990be}, a~general parametrization of the dark halo 
in galaxies can written as (reference~\cite{Hernquist:1990be} has nothing to do with dark matter, but~on the density distribution of visible mass in
  spherical galaxies. Equation~(\ref{eq:profiles}) here is an~adaptation of Equation~(43) in that paper).

  \begin{equation}
\rho_{DM}(r)\,=\,\frac{\rho_0}{\left ( \delta + \frac{r}{r_s} \right )^\gamma \cdot \left ( 1 + (\frac{r}{r_s})^\alpha   \right )^{(\beta-\gamma)/\alpha} }
\label{eq:profiles}
\end{equation}
where $r$ is the distance from the Galactic Centre, the~parameters $\alpha$, $\beta$ and $\gamma$ determine the shape of the halo and $\rho_0$ is a~normalization constant
depending on the model, i.e.,~depending on the choice of $\alpha$, $\beta$ and $\gamma$. Different combinations of these parameters can recover the  halo profiles
mentioned above.  
These profiles describe the smooth distribution of dark matter around galaxies. Any substructure (clumpiness)  must be added on top in order to describe the outcome of N-body
simulations~\cite{Taylor:2002zd,Ando:2019xlm}.  The~effect of clumpiness is usually parametrized as a~multiplicative boost factor to the expected signal
from annihilations without substructure~\cite{Moline:2016pbm,Penarrubia:2017nzw}  and it is, by~necessity, an~average of the enhancement given by simulations of a~population of~halos.

Figure~\ref{fig:DMprofiles} shows that the different halo models proposed show agreement on the dark mater distribution at the location of the Solar 
System (about 8 kpc from the Galactic Centre) and beyond, but~the predicted dark matter density distribution diverges considerably at distances near 
the centre of the galaxy~\cite{Weber:2009pt}. This is of course also true when considering other galaxies and assuming the same type of profiles apply universally at any redshift.
However, universality might not be completely justified, and~the shape of a~dark halo can depend on local characteristics of any given galaxy. There~are indications that halo profiles
might depend both on the size of the galaxy and the structure of its surroundings, since the original mass function of nearby satellites will affect the accretion of additional dark matter to
the halo, causing the inner slope of the dark matter profile to deviate form a~universal function like the ones in Equation~(\ref{eq:profiles})~\cite{Ricotti:2002qu}. Indeed, recent results from
the Gaia mission point to a~more complex dark matter halo in the Milky Way than just a~spherical distribution. Gaia will map the position and radial velocity of about one billion stars,
providing and unprecedented 3-dimensional picture of our galaxy and invaluable data to precisely map the dark matter halo for the first time.  
Gaia's data release 2, along with data from the Sloan Digital Sky Survey (SDSS), have already provided indications
that our dark matter halo has a~non-trivial structure, with~localized substructures due to the recent history of our Galaxy that might not yet be in equilibrium. Actually, the~Gaia
results seem to be in tension with the characteristics of the dark halo derived from previous stellar data~\cite{Cautun:2019eaf,Buch:2018qdr,Necib:2018iwb,OHare:2019qxc,Banik:2019cza}. 
The first implication of these results is that the Milky Way halo is more complex than initially thought, with~local substructure that now can be mapped. This can have implications for
direct dark matter searches whose expected signal depends on the assumed dark matter density in the solar~neighborhood.

However, in the bigger picture, the~Gaia results point to a~dark matter halo which
is strongly dependent on the specific history of the galaxy, determined by past accretion events from nearby satellite galaxies.  If~dark matter halos strongly depend on the individual
history of each galaxy, the~assumption of universality of the dark matter profiles breaks down, with~consequences for the interpretation of indirect dark matter searches as well.
In particular there are two related aspects to take into account: The integrated rate of supernova explosions through the evolutionary history of a~galaxy and the role of baryon feedback
in the star formation cycle of the galaxy~\cite{Dekek:1986gu,Ogiya:2012jq,Weinberg:2013aya,DelPopolo:2015nda,Salucci:2018hqu}. The~ejecta from supernova explosions can heat the surrounding
interstellar gas, injecting energy into it to a~point that further star formation due to gravitational collapse significantly slows down, or~even stops. As~the affected gas expands and cools down, gravity draws it back
towards the center of the galaxy, reinitiating the star formation process and eventually leading to more supernovae explosions. For~large galaxies, with~a large amount of gas, the~cycle
from star formation to explosion to re-ignition of star formation can be quite fast. During~these processes the baryonic mass distribution of the galaxy fluctuates, inducing a~change
in the gravitational potential of the gas which can have an~effect on the dark matter distribution. In~particular, this process could solve the core-cusp problem, since the net effect
on the dark matter component is to redistribute the dark matter away from the galactic center, converting an~initially cusped profile in a~cored profile. The~rate of supernova ejecta and
the strength of baryon feedback, and~their effect on the dark matter distribution of a~galaxy, depend on the size, age, evolution~and surroundings of the galaxy, which support the idea
that universal halo functions are just a~first approximation to the structure of a~galaxy, but~that individual characteristics need to be taken into account. This can affect the calculations
of $J$-factors discussed~below.

Barring the caveats just mentioned, given a~specific dark matter halo, the~flux of cosmic-rays, gamma-rays or neutrinos arising from dark matter annihilations in a~given object can
be expressed as
\begin{equation}
 \frac{dN_i}{dA dt dE_i}\,=\,\frac{d\Phi_i}{dE_i}\,=\,\frac{1}{4\pi}\,\frac{\left < \sigma_{\rm A} v \right >}{2 m^2_{\rm DM}}\,\frac{dN_i}{dE_i} \times \int_{\Omega} \int_{l.o.s.} \rho^2_{DM}(r) dr d\Omega
\label{eq:indirect_galaxy}
\end{equation}
where $i$ stands for the type of particle, $\left < \sigma_{\rm A} v \right >$, is the thermally averaged product of the dark matter self-annihilation cross-section 
times the dark matter velocity, $dN_i/dE_i$ is the spectrum of species $i$ produced by the annihilations, $\rho_{DM}$ is the dark matter density of the source, given 
by Equation~(\ref{eq:profiles}) and $m_{\rm DM}$ is the dark matter mass. The~integral of the squared density along the line of sight (l.o.s.) is the so-called $J$-factor. As~argued above,
the $J$-factor is source-specific, and~encompasses our knowledge of the structure of the dark halo of the source in consideration. The~$J$-factor depends on the 
halo profile chosen as well as on the distance to the source, and~results from indirect dark matter searches must be given under the assumption of a~specific halo model. Clearly, 
distances need to be estimated independently, which for far away objects can add to the final uncertainty of the result. 
$J$-factors are  measured in $GeV^2 cm^{-5} sr$.
As Figure~\ref{fig:DMprofiles} illustrates, using one parametrization of $\rho_{DM}$ or another can give results that differ by as much as orders of~magnitude.

Assuming a~particle physics model which gives the expected particle spectrum, $dN_i/dE_i$, and~a~halo model, then an~experimental 
measurement of $d\Phi_i/dE_i$ can be used to probe the two independent terms in Equation~(\ref{eq:indirect_galaxy}): $\left < \sigma_{\rm A} \textrm{v} \right >$
versus the dark matter mass, $m_{\rm DM}$. Experimental efforts are usually kept as model-independent as possible, looking for signatures from generic candidates
over a~few benchmark masses that cover a~wide mass range and, since the exact branching ratios of dark matter annihilations into different channels
are unknown, analyses are typically optimized assuming 100\% annihilation to a~few characteristic channels. This brackets the expected spectrum from any
realistic model which would produce an~spectrum that is a~weighted mixture of all the possible annihilation~channels.

If signals from dark matter decay, instead of annihilation, are to be probed, then Equation~(\ref{eq:indirect_galaxy}) needs to be slightly modified: 
The l.o.s. integral is over $\rho$ and not $\rho^2$ (only one particle participates in a~decay, while two are needed in an~annihilation), only
one $m_{\rm DM}$ factor is needed in the denominator (for the same reason), and~what is probed is $1/\tau$ instead of $\left < \sigma_{\rm A} v \right >$,
where $\tau$ is the candidate lifetime. In~this case the units of the $J$-factor are  $GeV cm^{-2} sr$.

It is worth pointing out that the term $\left < \sigma_{\rm A} v \right >$ can provide more information than it might seem at first sight. It is not only sensitive to
a given particle model, through what the model predicts for $\sigma_{\rm A}$, but~it is also sensitive to the structure of the host galaxy (or the medium where the annihilations take place), 
through the velocity distribution from which the annihilating particles are drawn. For~2$\rightarrow$2 annihilations in the non-relativistic case, the~term $(\sigma_{\rm A} v)$ can be expanded
in powers of angular momentum in the square of the relative velocity of the particles as $\sigma_{\rm A} v~=~\sigma_s + \sigma_p v^2 +...$, where $\sigma_s$ and $\sigma_p$ are constants
(see~\cite{Gondolo:1990dk} for the corresponding relativistic expression and a~thorough discussion on the applicability regimes of $\left < \sigma_{\rm A} v \right >$). 
The previous expression explicitly shows the constant behaviour of s-wave ({\em{l}}~=~0)  scattering and the quadratic dependence on the relative velocity of p-wave ({\em{l}}~=~1) scattering.
Given the dark matter velocity in the halos of galaxies, $\beta~\sim 10^{-6}$, the~relative velocity of two annihilating particles is typically low enough to make the
p-wave annihilation rate subdominant to the s-wave term. However, s-wave annihilations to some channels (fermions for example) can be chirality suppressed by a~factor $m_f/m_{\rm DM}$,
where $m_f$ is the fermion mass, so the prospects of indirect detection of certain dark matter candidates can become slim.
There are, however, natural ways out of this problem that bring back the hope for some of the indirect detection channels.
One is simply to search for dark matter annihilations in objects which are violent enough for dark matter to orbit with high velocities, where p-wave annihilations naturally
dominate~\cite{Johnson:2019hsm}. Further, if~one considers annihilations with three-body final states, 2$\rightarrow$3, where the third particle arises as a~$\gamma$, Z or W bremsstrahlung
in the final state, the~helicity suppression is highly ameliorated. Besides, these type of annihilations can simultaneously produce photons and electrons, or~photons and neutrinos,
providing a~multi-messenger signal from the same process. An~interesting study case for experimentalists. Another way to avoid the suppression of s-wave annihilations is through the
Sommerfeld enhancement in models with self-interactions in the dark matter sector~\cite{ArkaniHamed:2008qn,Feng:2010zp,Campbell:2010xc,Das:2016ced,Choi:2017mkk,Clark:2019nby}. In~such case
the new potential enhances the s-wave cross section through the well known Sommerfeld effect~\cite{Sommerfeld:1931aa} by factors as high as three orders of magnitude, making such models
testable with the messengers and sources mentioned in the sections~below.

There is a~further point to consider concerning the use of Equation~(\ref{eq:indirect_galaxy}). As~it has been presented it includes only the spectrum from the source under consideration but it
ignores the diffuse cosmological contribution from all the sources along the line of sight, at~different redshifts. This contribution can be sometimes safely neglected, for~example in
searches from nearby objects like the center of the Milky Way where the additional signal from the far cosmos is subdominant. However, it can be of
relevance when pointing at extended sources or in analyses with poor angular resolution. In~case this effect needs to be included, a~cosmological model needs to be assumed and energies
detected at the Earth need to be properly redshifted, and~an integral over redshift performed. These effects have been clearly described, for~example for gamma-rays and neutrinos,
in~\cite{Ullio:2002pj,Beacom:2006tt}. Diffuse cosmological contributions can also be important when considering halos with substructure. Since dark matter annihilations scale as the dark matter
density squared, high-redshift but clumpy halos could contribute a~non-negligible signal at~Earth.

\section{Dark matter signatures from the cosmos}
\label{sec:signatures}
Since the expected signal from dark matter annihilation (decay) is proportional to $\rho_{DM}^2$($\rho_{DM}$), natural sources to consider in indirect searches
are regions in the universe with an~increased density of dark matter. Typically,  our own Milky Way, close-by dwarf spheroid galaxies and galaxy clusters are
considered in indirect searches. Dwarf spheroid galaxies are good candidates because they present high mass-to-light ratios, with~little interstellar gas, and~they
seem therefore to be dominated by dark-matter. Additionally, there is no evidence of non-thermal emission from any of the 20+ known dwarf spheroids that 
could produce a~background of high energy neutrinos, $\gamma$-rays or cosmic-rays so they are a~clean target from the observational point of view. Even if they are
small systems, with~$J$-factors much smaller than our Galactic Centre, observations of several dwarves can be stacked in order to increase the sensitivity to a
potential dark matter~signal. 
 
Indirect searches for dark matter have indeed been performed by 
gamma-ray telescopes (MAGIC, e.g.,~\cite{Acciari:2020pno,Acciari:2018sjn,Giammaria:2016jfo}, H.E.S.S., e.g.,~\cite{Rinchiuso:2019etv,Armand:2019ive,Abdalla:2018mve,Abdalla:2016olq}, VERITAS~\cite{KhanCantlay:2019syo,Archambault:2017wyh}), X-ray telescopes (XMM-Newton, e.g.,~\cite{Boyarsky:2017wct,Ruchayskiy:2015onc,Borriello:2011un,Boyarsky:2007ay}, NuSTAR, e.g.,~\cite{Neronov:2016wdd,Perez:2016tcq,Roach:2019ctw}, Suzaku, e.g.,~\cite{Sekiya:2015jsa,Tamura:2014mta,Urban:2014yda,Kusenko:2012ch}), cosmic-ray detectors
(HAWC~\cite{Cadena:2019lor,Abeysekara:2017jxs,Albert:2017vtb}), detectors~in space (Fermi-LAT, e.g.,~\cite{Abazajian:2020tww,DiMauro:2019frs,Fermi-LAT:2016uux}, DAMPE, e.g.,~\cite{Wechakama:2019vzk}, CALET~\cite{Adriani:2018ktz}, AMS~\cite{Xu:2020zfh})
and neutrino telescopes (IceCube, e.g.,~\cite{Aartsen:2018mxl,Aartsen:2017ulx,Aartsen:2016zhm}, ANTARES, e.g.,~\cite{ANTARES:2019svn,Albert:2016dsy,Adrian-Martinez:2016gti}, Baikal, e.g.,~\cite{Avrorin:2016yhw,Avrorin:2015bct,Avrorin:2014swy}, Baksan~\cite{Boliev:2013ai} or Super-Kamiokande, e.g.,~\cite{Desai:2004pq,Abe:2020sbr,Choi:2015ara}). Many of these detectors have upgrade programs underway, in~different stages of R\&D or
implementation (CTA~\cite{Consortium:2010bc}, IceCube-Gen2~\cite{Aartsen:2020fgd}, KM3NET~\cite{Adrian-Martinez:2016fdl}, Baikal-GVD~\cite{Avrorin:2019dov}, Hyper-Kamiokande~\cite{Abe:2011ts}),
which will bring indirect dark matter searches to the next level in mass coverage and~sensitivity. 

Each messenger has its quirks and features, though, both from the point of view of propagation and/or detection. 
In order to increase the sensitivity to specific sources that can be observed with different instruments, several collaborations have joined forces and obtained limits from
their combined observations. Combining results from different experiments requires a~consensus on the use of common $J$-factors for each target, statistical techniques and
the treatment of the systematic uncertainties of the different instruments. The~result of these efforts are usually stronger limits than those obtained by the individual
collaborations. They also help unify analysis methods and the use of common external inputs (astrophysical quantities, cross sections, dark matter models...) across the
community, which in the long run helps obtaining a~coherent view of the field and makes it easier to compare, and~even to combine, results from different messengers.
One can really talk of a~multi-messenger era in dark matter~searches. 

\subsection{Gamma- and~X-rays}
\begin{figure}[t]
\centering
\includegraphics[width=0.45\linewidth,height=0.40\linewidth]{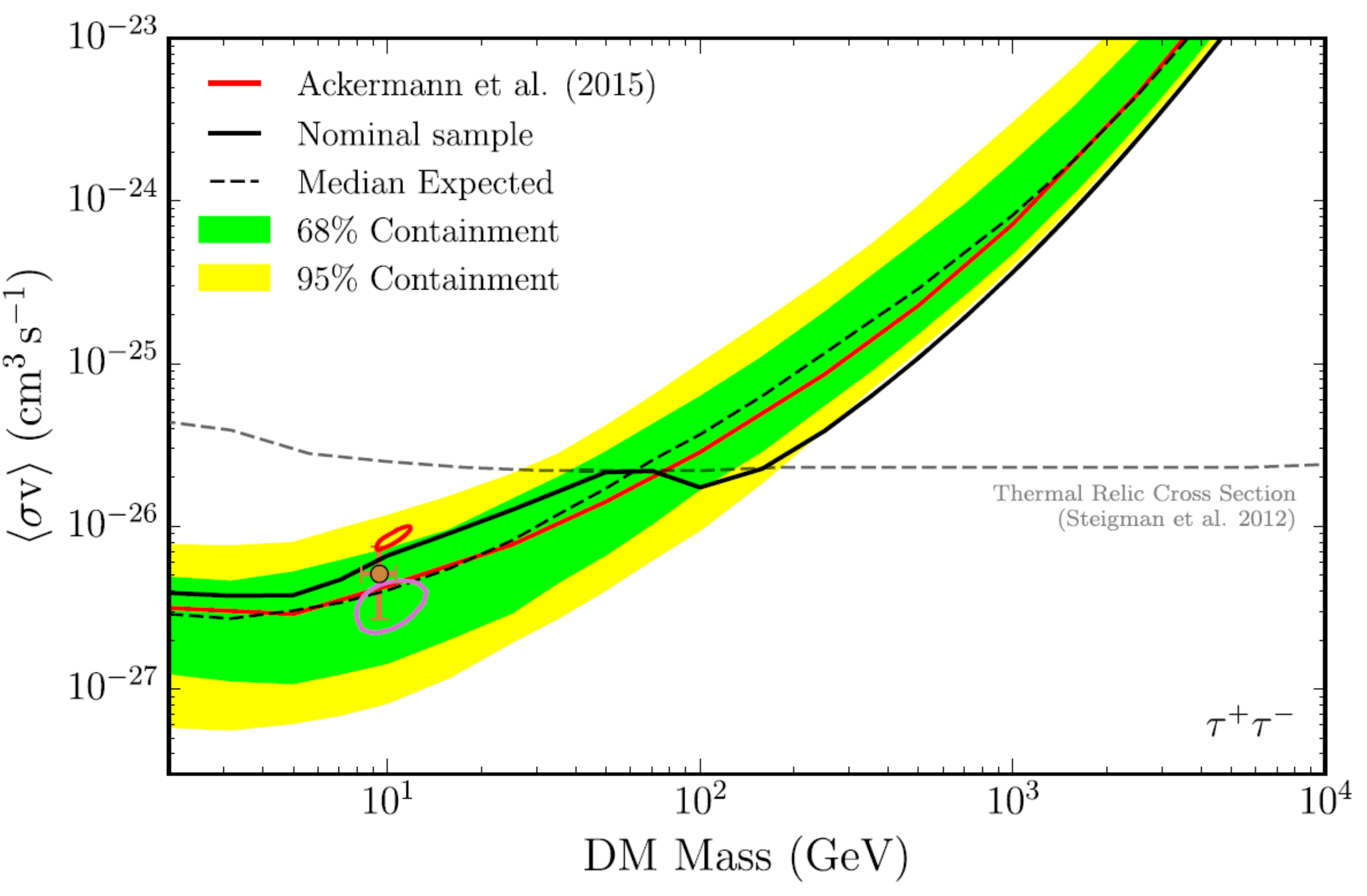} 
\includegraphics[width=0.43\linewidth]{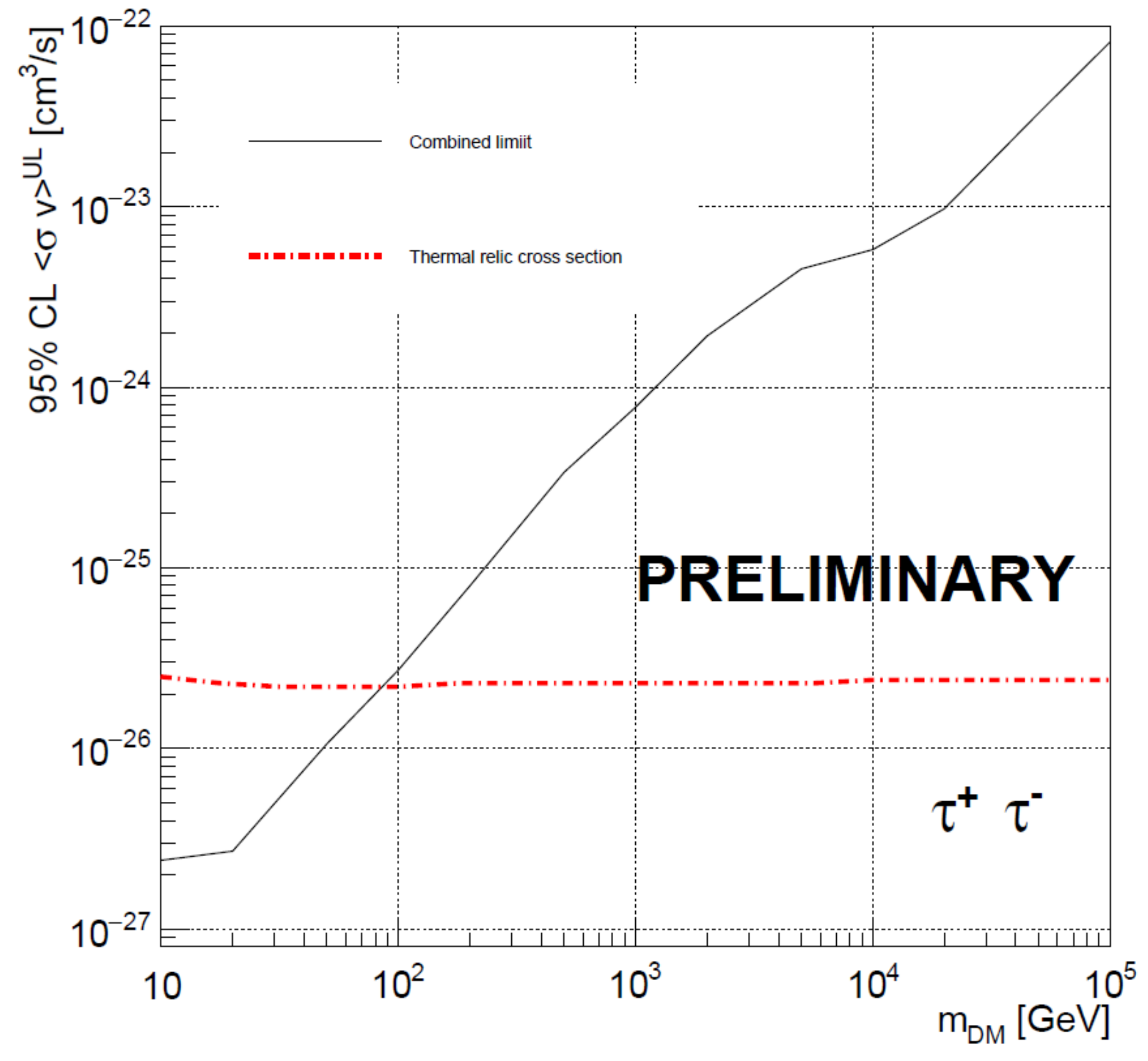}
\caption{ (\textbf{Left}) The 95\% confidence level upper limits  on the velocity-averaged dark matter annihilation cross section obtained from 45 dwarf galaxies with the Fermi-LAT instrument, assuming 
  100\% annihilation to $\tau^+\tau^-$. The~yellow and green lines represent the 95\% and 68\% quantiles respectively. The~solid red line shows
the limit from a~previous analysis using 15 sources. The~closed contours and marker show the best-fit regions (at 2$\sigma$ confidence) of several dark matter interpretations of the Galactic Centre excess.  Figure reprinted with permission from~\cite{Fermi-LAT:2016uux}. Copyright 2020 AAS.  
(\textbf{Right}) The 95\% confidence level upper limits on the dark matter annihilation cross from the combination of data from observations of
20 dwarf galaxies by Fermi-LAT, HAWC, H.E.S.S., MAGIC and VERITAS,  assuming  100\% annihilation to $\tau^+\tau^-$. Figure from~\cite{Oakes:2019ywx}. } 
\label{fig:gamma-dwarves}
\end{figure}

Gamma-rays propagate without deflection, so they point to their source, and~are relatively easy to detect, the~technology being available since long. 
Gamma-ray telescopes and gamma-ray detectors on space are able to point with great accuracy (about 0.1 degree for single photons above 10 GeV in Fermi-LAT) and
the same for Cherenkov telescopes like MAGIC or H.E.S.S. for example (although at higher energies, above~about 100~GeV) and they
cover a~wide range of photon energies, from~20~GeV in Fermi-LAT to few hundred TeV in Cherenkov telescopes. However, absorption and backgrounds are always a~concern when searching
for new physics with photons from space. The~Milky Way centre is a~quite crowded region with $\gamma$-emitting objects and a~diffuse component from unresolved sources and star
formation regions. In~this respect, the~clean structure of dwarf spheroid galaxies is expected to provide a~favourable environment for dark matter searches with gammas, and~indeed strong limits on the dark matter annihilation cross section have been obtained from these objects. An~illustration of the results that can be obtained with this kind of 
searches is shown in the left plot of Figure~\ref{fig:gamma-dwarves}. The~figure shows 95\% confidence level upper the limits on the velocity-averaged dark matter annihilation cross section
as a~function of dark matter mass obtained from observations of 45 dwarf galaxies with the Fermi-LAT telescope~\cite{Fermi-LAT:2016uux}. Refined limits can be obtained exploiting the different
energy reach (and~therefore dark matter mass sensitivity) of different instruments, as~it is shown in the right plot of Figure~\ref{fig:gamma-dwarves}. The~plot shows a~combined limit
on $\left < \sigma_{\rm A} v \right >$ using results from Fermi-LAT, HAWC, H.E.S.S., MAGIC~and VERITAS, and~reaches an~improvement of up to one order of magnitude for masses above 10~TeV due to
the contribution of the Cherenkov telescopes~\cite{Oakes:2019ywx}.

The uncontroversial negative results from dwarf galaxies can be confronted with the status of dark matter searches with gammas from the Galactic Centre. 
An intriguing excess of gamma-rays in the few GeV region over the background expected from conventional sources has been claimed by several authors, both from EGRET
data~\cite{Hunger:1997we} and more recently from Fermi-LAT data~\cite{Goodenough:2009gk}. However, an excess is of course defined with respect to a~background, and~here is where difficulties
arise. The~region around the Galactic Centre is one of the more complex in the sky, not least in gamma-rays. It is a~star-forming region, surrounded with interstellar gas
and with plenty of individual sources, like short-period pulsars and supernova remnants, as~well as allegedly unresolved sources. Gamma-rays can be produced in point sources but
also as a~diffuse glow  by electron bremsstrahlung, by~inverse Compton scattering of electrons and positrons on the intense ambient radiation field in the region, or~as products of
the interactions of cosmic-rays with the interstellar gas. The~calculation of the expected gamma-ray flux from all these processes, along with the galactic model used to track the
diffusion of cosmic-rays in their way through the galaxy, is plagued with uncertainties which reflect in the predicted gamma ray emission from the Galactic Centre (which is the
background when searching for dark matter)~\cite{Calore:2014xka,TheFermi-LAT:2017vmf,Buschmann:2020adf,Zhong:2019ycb}. Indeed the EGRET excess was promptly explained by detailed
calculations of the expected cosmic-ray flux without the need of new physics~\cite{Bergstrom:2006tk}. An~additional important uncertainty in modeling the inner region of the Galaxy
arises from the ``Fermi bubbles''~\cite{Su:2010qj}. Their contribution to the gamma ray glow is difficult to estimate since information is lacking about their structure and emitted
spectrum close to the Galactic Centre. Masking known point sources and subtracting the diffuse emission from the Galactic Centre in order to characterize the excess is therefore a
really complicated task, plagued with unknowns. However, a new method of analyzing photon sky maps, non-Poisson template fitting, has been proposed in~\cite{Lee:2015fea} and applied to the
inner galactic region. The~method is more sensitive to unresolved point sources than traditional Poisson-count based analyses, and~shows that the Fermi-LAT gamma-ray excess can be explained
by the background emission produced by a~population of sub-threshold objects, without~the need of dark~matter.

However, the location of the excess, within~1.5$^\circ$ of the Galactic Centre, is indeed a~place where an~increased density of dark matter is expected. Given the unknowns in any dark
matter model, the~excess can of course be also fitted as originating from dark matter annihilations, as~indeed it has (see, e.g.,~\cite{Hooper:2010mq,Hooper:2011ti,Agrawal:2014oha,Calore:2014nla,Achterberg:2017emt}
although the literature is vast on this topic).  
The main ``result'' of these claims is that a~dark matter particle 
in the mass range of a~few tens of GeV to a~few hundred GeV, annihilating~to soft (typically $b\bar{b}$ or $\tau^+\tau^-$) or hard (typically $W^+W^-$ or $t\bar{t}$), channels
can fit the data. The~wide range of masses and annihilation channels proposed simply reflects the underlying uncertainties in estimating the background, as~well as the rather 
unconstrained options that the dark matter solution provides. The~allowed parameter space in 
dark matter models is ample enough to be able to fit the Fermi-LAT gamma ray excess without problem, even with the oversimplifying assumption of a~single
one-particle species annihilating through just one channel and taking into account limits from new particle searches in~accelerators.  

\begin{figure}[t]
\centering
\includegraphics[width=0.70\linewidth]{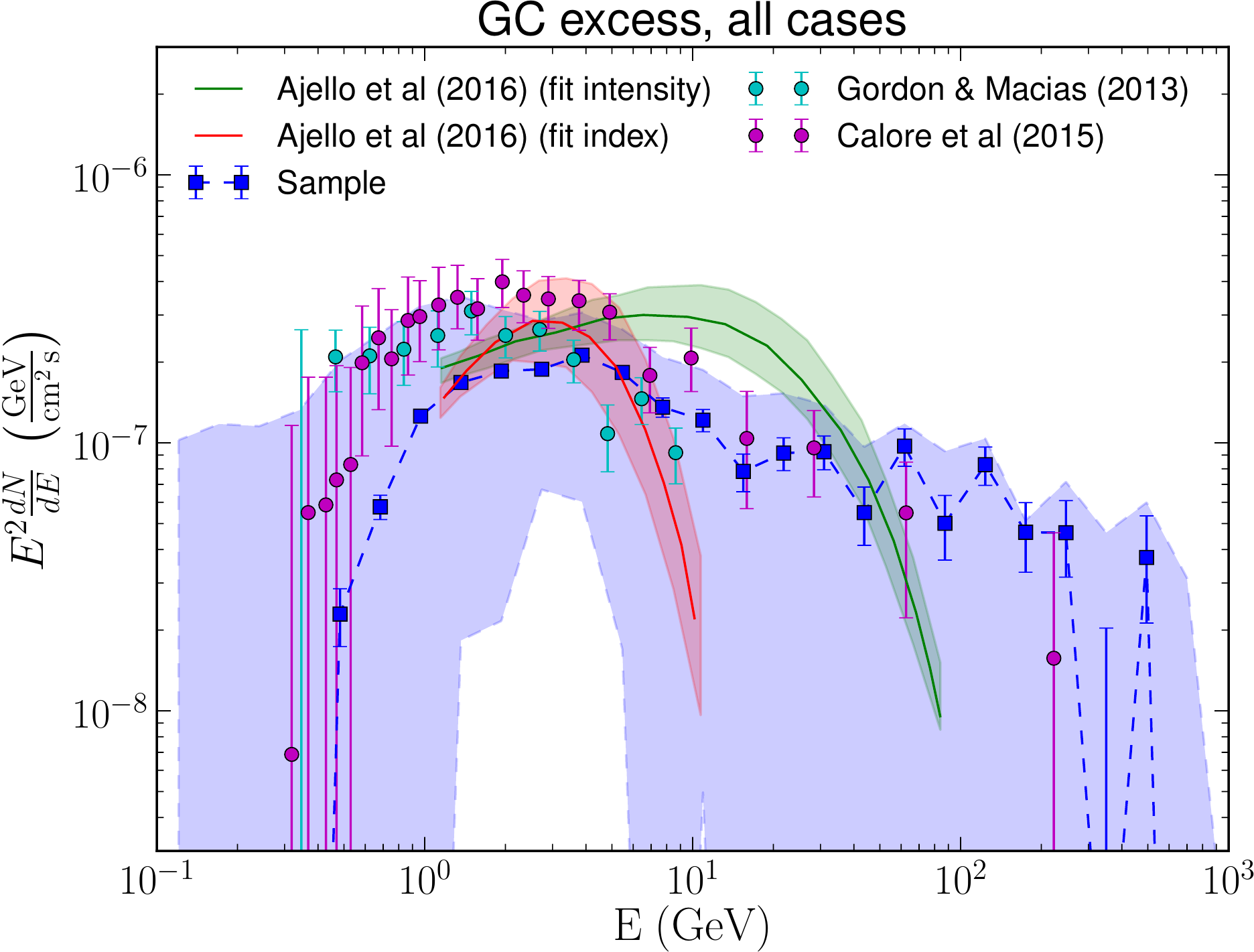} 
\caption{ Spectrum of the Galactic Centre excess. The~systematic uncertainty band shows the envelope of different Galactic Centre excess fluxes obtained using different
  models of diffuse gamma-ray emission and source masking. Figure reprinted with permission from~\cite{TheFermi-LAT:2017vmf}. Copyright 2020 AAS. See reference for details of the analyses
mentioned in the inset caption.}
\label{fig:Fermi_GCexcess}
\end{figure}

The authors in~\cite{TheFermi-LAT:2017vmf} have performed an~extensive study of the Galactic Centre and used several models of the galaxy to estimate the expected
gamma ray glow, finding that an~excess over the expected gamma ray flux from the Galactic Centre can be obtained in all the models considered. However, they also point out that the
excess is still compatible, given the huge uncertainties, with~the emission from standard astrophysical sources when considering contributions from cosmic-ray interactions in the
interstellar medium or unresolved populations of pulsars. It is, of~course, also compatible with ad-hoc dark matter models but, again, Occam's razor renders at this point the dark
matter interpretation of the gamma ray excess from the Galactic Centre as, at~most, a~mere viable solution, but~not the only one. A~summary of this study, in~comparison with the original
claims of the excess, is shown in Figure~\ref{fig:Fermi_GCexcess}.
The~figure shows the expected emission from the Galactic Centre obtained by the Fermi-LAT collaboration with a~detailed modeling of the Galaxy (points tagged as ``Sample'') compared with the
excess claimed by other authors, using different assumptions on the morphology of the Galaxy and background templates (so, in~a strict sense, the~curves are not really directly comparable).
The~shaded area represents the systematic uncertainty from the Fermi-LAT collaboration analysis, and~it is obtained as the envelope of the predictions of the gamma-ray flux under different
assumptions on diffuse gamma-ray emission, galaxy morphology, gas distributions in the galaxy or cosmic-ray distribution and propagation, but, in~any case, without~any new physics.
A similar conclusion is reached in~\cite{Abazajian:2020tww}, where diffuse galactic emission, inverse-Compton emission and a~central source of electrons can explain the Fermi-LAT data without
a dark matter component, practically ruling out thermal dark matter as the explanation of the galactic centre gamma-ray~excess.

\begin{figure}[t]
\centering
\includegraphics[width=0.45\linewidth,height=0.40\linewidth]{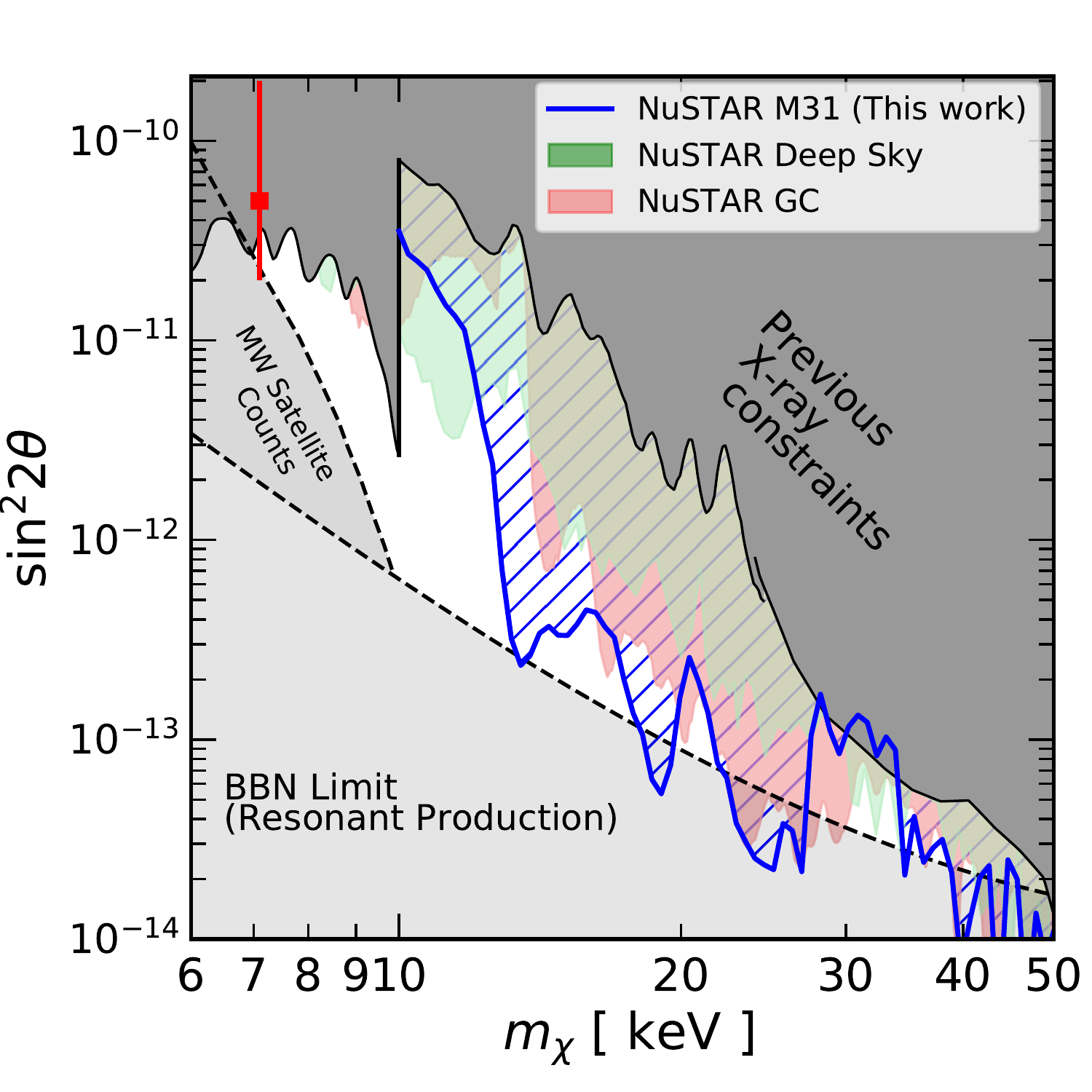} 
\includegraphics[width=0.43\linewidth,height=0.40\linewidth]{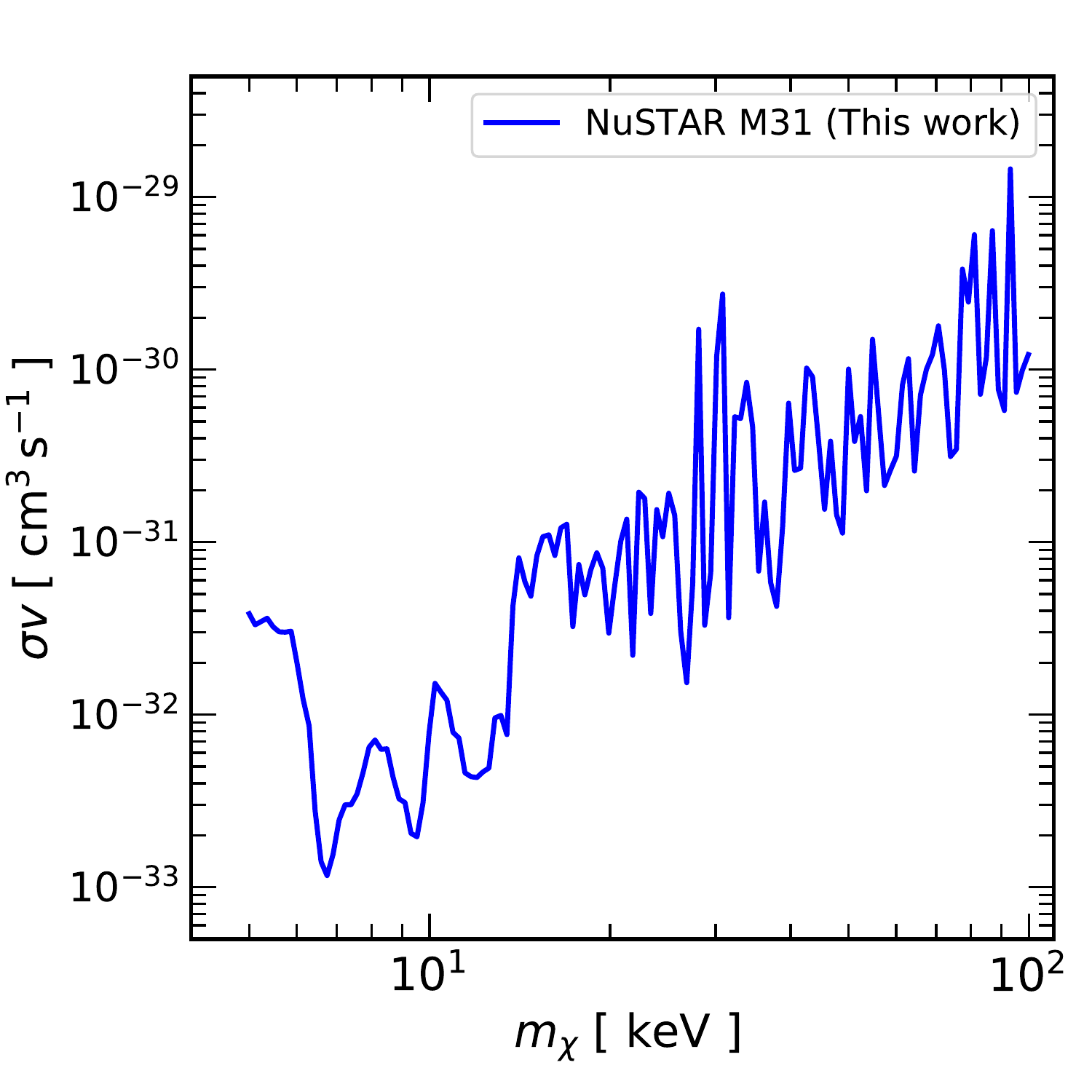}
\caption{ (\textbf{Left}) Excluded areas of the sterile neutrino mass and mixing angle parameter space from NuSTAR observations of M31, the~Galactic Center and
  the deep X-ray sky. See reference below for details.  
  (\textbf{Right}) Upper limit on weakly interacting dark matter annihilation cross section from NuSTAR observations of M31, assuming annihilation to two monoenergetic photons. 
 Figures reprinted with permission from~\cite{Ng:2019gch}. Copyright 2020 by the American Physical Society.
}
\label{fig:xray-limits}
\end{figure}

While the rather energetic gamma-rays can probe the $(\left < \sigma_{\rm A} v \right >, m_{\rm DM})$ parameter space for dark matter masses of $\mathcal{O}$(few GeV) and above,
the softer diffuse X-ray sky can be used to probe dark matter masses down to the keV region~\cite{Essig:2013goa,Jeltema:2011bd}. X-rays can be produced directly from low mass dark matter annihilations or decays,
but they can also be produced by inverse Compton scattering of $e^+e^-$ from the annihilations or decays of dark matter of any mass on background photons. Inverse Compton scattering
of charged annihilation or decay products on CMB photons, dust or starlight will produce a~diffuse photon flux with typical photon energies  of the order of
1--100 ($m_{\rm DM}$/10~GeV)$^2$~keV~\cite{Jeltema:2011bd}, where $m_{\rm DM}$ is the dark matter mass, and~the energy range covers therefore the X-ray region. X-ray~telescopes
provide, therefore, a~complementary way to search for dark matter from galaxy clusters and galaxies, including the Milky Way. One of the best theoretically justified light dark matter candidate
in the keV mass region is a~right-handed (i.e., sterile) neutrino, which can 
be accommodated in extensions of the standard model without many additions to the particle spectrum of the theory~\cite{Dodelson:1993je,Adhikari:2016bei,Boyarsky:2018tvu,Abazajian:2017tcc}.
Sterile~neutrinos can be produced in the early universe through mixing with active neutrinos, so the mass  $M_s$ of the sterile neutrino and the mixing angle with
active neutrinos are two fundamental parameters of any model. Probing the $(M_s,{\rm sin}^2(2\theta))$ parameter space amounts to probing the feasibility of sterile neutrinos
to be dark matter. Sterile neutrinos have a~radiative decay channel into an~active neutrino and a~photon ($\nu_s \rightarrow \nu \gamma$). The~integrated effect of decays at different redshifts during the evolution of the universe would result in a~broad X-ray band today on top of the 
the diffuse X-ray astrophysical background. The~non-detection of such feature has already allowed to constrain the $(M_s,{\rm sin}^2(2\theta))$ parameter space of sterile neutrino dark
matter with X-ray telescopes~\cite{Roach:2019ctw,Ng:2019gch,Boyarsky:2017wct,Ruchayskiy:2015onc,Sekiya:2015jsa,Tamura:2014mta,Urban:2014yda,Borriello:2011un,Abazajian:2006jc,Boyarsky:2005us}. 
The left plot of Figure~\ref{fig:xray-limits} shows that X-ray telescopes alone have a~strong constraining power on 
the $(M_s,{\rm sin}^2(2\theta))$ plane, in~this case illustrated on the $\nu$MSSM model, a~minimal extension of the 
standard model that includes neutrino masses and which incorporates a~dark matter candidate in the form of a~right-handed neutrino of a~few keV~\cite{Asaka:2005an}.
Additionally, the~same type of analyses can be used to constrain the annihilation cross section of a~generic low-mass dark matter candidate. The~right plot of Figure~\ref{fig:xray-limits}
shows the limit on the velocity averaged annihilation cross section of a~generic light weakly interacting dark matter candidate assuming annihilation to two photons obtained with 
the same X-ray observations as in the left plot. While~this result has been obtained for a~quite favourable situation for X-ray detectors, direct annihilation into photons, the~
limit on  $\left < \sigma_{\rm A} v \right >$ is quite strong if compared with the results from other~messengers.

\unskip 

\subsection{Neutrinos}
Neutrinos also propagate undeflected and therefore point to their source, but~they present additional challenges compared with photons. First, their extremely low cross section with matter
means that large detectors and exposure times are needed to collect just an~event from a~given source, not to mention a~handful of events that would be statistically significant by themselves.
In this sense, the~best bet of neutrino telescopes to function as dark matter detectors is to correlate their findings with other messenger or, ideally, more than one. This would, in~turn,
help interpreting any signal from gamma-rays or cosmic-rays, since an~excess of events from a~single messenger might be subject to interpretation as background from 
other processes rather than from dark matter, or~it would need quite ad-hoc cooked theoretical models to explain why the dark sector annihilates or decays to some standard model 
particles but not others. 
Besides, neutrino detectors have a~limited angular resolution, of~the order of 1$^o$ at $\mathcal{O}$(100~GeV) neutrino energy, only reaching below one degree at TeV energies, and~therefore 
pointing becomes challenging in searches for relatively low-mass dark matter candidates. Even in a~perfect detector the angular resolution would be ultimately limited by the physics of the
neutrino-nucleon cross section. Limited pointing resolution impacts the degree of background rejection, mainly atmospheric neutrinos,
when pointing towards a~source. On~the other hand, neutrinos have the advantage of not loosing energy or being absorbed through their propagation over cosmic distances. Neither there is a
significant background or foreground from astrophysical objects, so they remain an~attractive signature. 
They are indeed the only possible messengers to use when looking for dark matter accumulated in dense objects like the Sun or the Earth. More on this~below.

 The total number of signal events, $\mu_{\rm s}$, expected at a~neutrino telescope is given by,
\begin{equation}
 \mu_{\rm s} =  T_{\rm live}\sum_\alpha\int{\rm d}\Omega\int {\rm d} E_\nu A^{\rm eff}_{\nu_\alpha}(\Omega,E_\nu) \Phi_{\nu_\alpha}(E_\nu)\,,
\label{eq:dm_nevents}
\end{equation}
where $T_{\rm live}$ is the exposure time, $A^{\rm eff}_{\nu_\alpha}(\Omega,E_\nu)$ the detector effective area for neutrino flavour $\alpha$, that~depends on the detector
response with respect to the observation angle and neutrino energy and $\Phi_{\nu_\alpha}(E_\nu)$ is the neutrino flux for flavour $\alpha$ arising the annihilation processes.
Note that the effective area of neutrino telescopes is not a~fixed quantity, but~it depends not only on neutrino energy and arrival direction, but~also on the specific
event selection performed. In~the absence of a~signal, the~90\% confidence level limit on the number of signal events, $\mu_{\rm s}^{\rm 90}$, can be directly translated into a~limit
on the neutrino flux and, in~turn, into~a limit on either the velocity-averaged annihilation cross section $\langle \sigma_{\rm A}\,v \rangle$ or the dark matter lifetime as a~function
of dark matter mass, through Equation~(\ref{eq:indirect_galaxy}). As~in the case with gamma-rays, sources with similar characteristics, where the signal can be expected to be also similar,
can be stacked to increase the signal-to-noise ratio. This is typically done when considering nearby dwarf~galaxies.

\begin{figure}[t]
\centering
\includegraphics[width=1\linewidth]{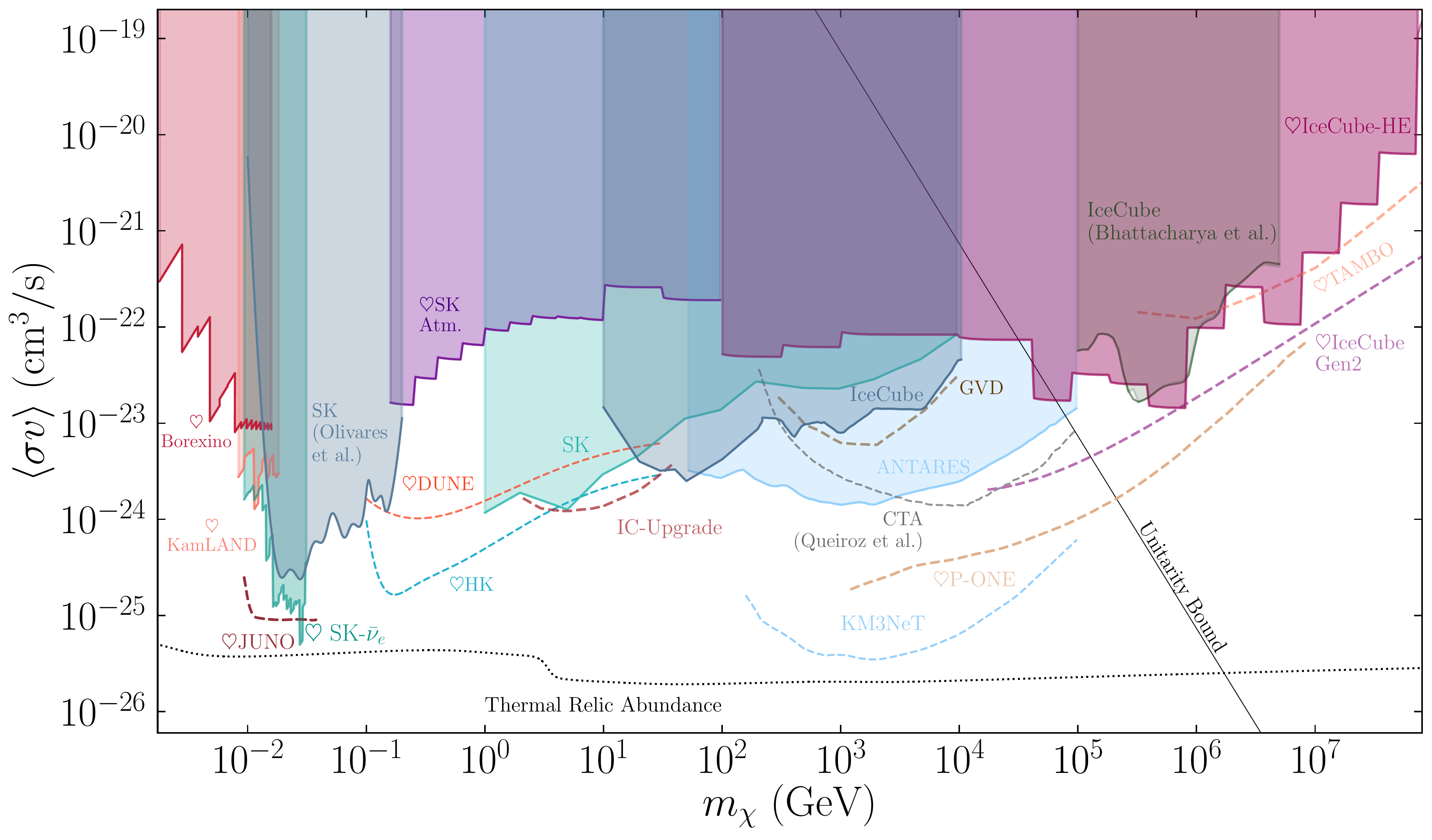} 
\caption{Summary of results on the velocity weighted  dark matter annihilation cross section from different experiments. Solid lines show limits, dashed lines sensitivities of
  future facilities assuming five years data taking (100 h of observation for the CTA sensitivity). The~heart symbols represent analyses performed by the authors
  of~\cite{Arguelles:2019ouk} with public data, and~not by the collaborations. Figure~from~\cite{Arguelles:2019ouk} }
\label{fig:neutrinos_summary}
\end{figure}

The current status of searches for dark matter with neutrinos from galaxies span a~large chunk of dark matter masses as illustrated in Figure~\ref{fig:neutrinos_summary}.
The figure  shows the current landscape of limits on the dark matter self-annihilation cross section as a~function of dark matter mass obtained by the main players in the neutrino
dark matter search industry, including many low-energy experiments which are not the focus of this review~\cite{Arguelles:2019ouk}. Next-generation experiments are shown as dashed-lines,
while~the shaded areas show current limits, some of them obtained independently from the collaborations by the authors (marked with a~heart). Results cover an~impressive eleven orders
of magnitude in dark matter mass, extending beyond the theoretical limit for a~thermal relic, where~results should be interpreted under different class of models than the vanilla WIMP
scenario (see e.g.,~\cite{Palomares-Ruiz:2020ytu} and references therein). 

In addition to our Galaxy and other objects at cosmological distances, the~Sun and the centre of the Earth are also two possible sources of dark matter annihilations that could be
detectable due to their proximity. The~effect of the Sun gravitational attraction over the lifetime of the Solar System can result in the capture dark matter from the local halo
through repeated scatterings of dark matter particles in orbits crossing the star~\cite{Press:1985ug,Krauss:1985aaa,Srednicki:1986vj,Gaisser:1986ha,Ritz:1987mh}. The~same applies
for the Earth~\cite{Gould:1987ir,Gould:1988eq}. The~only caveat is that only neutrinos from dark matter annihilations in the centre
of these objects can escape their dense interior. So it is only neutrino telescopes that can pursue dark matter searches from the centre of the Sun and Earth (with some exceptions
mentioned below).

The  expected neutrino flux from dark matter annihilations inside the Sun depends, among~other factors, on~the capture rate of dark  matter (which is proportional to the dark
matter-nucleon cross section), and~the annihilation rate (which is proportional to the velocity-averaged dark matter annihilation cross section). The~evolution of the number density
 of dark matter particles, $N_{\rm DM}$,  accumulated in the Sun interior follows the equation
\begin{equation}
  \frac{dN_{\rm DM}}{dt} = \Gamma_{\rm C} - 2\langle\sigma_{\rm A} v\rangle(N_{\rm DM}^2/2)\,,
  \label{eq:density}
\end{equation}
where $\Gamma_{\rm C}$ is the capture rate per unit volume. The~numerical factors take into account that annihilations consume two particles per interaction but
there are only $1/2$ of the particles available to form pairs. We have neglected a~possible ``evaporation'' term, proportional to $N_{\rm DM}$, which is only relevant for very small
dark matter masses, $\lesssim$ 4 GeV~\cite{Steigman:1997vs,Spergel:1984re,Gaisser:1986ha}, below~the range currently probed by neutrino
telescopes due to their relatively high energy threshold (precisely a~few GeV), but~that can be an~issue to take into account in future low-energy extensions of current projects.
Equation~(\ref{eq:density}) has an~equilibrium solution given by
\begin{equation}
  N_{\rm DM, eq}=\sqrt{\frac{\Gamma_{\rm C}}{\langle\sigma_{\rm A} v\rangle}}\, ,
 \label{eq:equilibriumDM}
\end{equation}
which represents the steady amount of dark matter accumulated in the Sun if capture and annihilation have reached equilibrium. Since the Sun is around 4.6~Gyr old, it is usually
assumed that equilibrium has indeed been achieved. In~this case, the~neutrino emission only depends on the capture rate, which~is directly related to the dark matter-nucleon scattering
cross section, $\sigma_{\rm{DM-N}}$, and~this allows to set limits on $\sigma_{\rm{DM-N}}$ from the measured (or rather, lack of) anomalous high-energy neutrino flux from the Sun. 
In models where equilibrium might not have been achieved, then an~assumption on the dark matter self annihilation cross section is necessary to derive predictions on the
neutrino signal from dark matter~annihilations.

The dark matter-nucleon cross section can be expressed in terms of a~spin-dependent, $\sigma_{\rm SD}$, and~spin-independent, $\sigma_{\rm SI}$, contributions~\cite{Engel:1992bf} and, 
since the Sun is mainly composed of Hydrogen (75\% of H and 24\% of He in mass)~\cite{Grevesse:1998bj}, the~capture of dark matter from the halo by the solar gravitational well
occurs mainly through the spin-dependent scattering. Heavier elements constitute less than 2\% of the mass of the Sun, but~can still play
a role when considering dark matter capture, since the spin-independent cross section is proportional to the square of the atomic mass number. These heavy elements
can also take part in the spin-dependent capture process if dark matter presents momentum-dependent interactions with normal matter, giving rise to an~increased capture rate in
comparison to previous calculations, that can have a~bearing in the interpretation of experimental results~\cite{Catena:2015uha}. Even if searches for dark matter with neutrinos
from the Sun are performed in the most model-independent way, there is of course the possibility to probe specific models, and~both the main collaborations or external authors
using public data have done so. These include limits on Kaluza-Klein dark matter arising in models of universal extra dimensions~\cite{Bernadich:2019zss},
inelastic dark matter~\cite{Catena:2018vzc}, strongly~interacting dark matter~\cite{Albuquerque:2010bt}, or~specific extensions of the
MSSM~\cite{Silverwood:2012tp,Trotta:2009gr,Allahverdi:2009se}, among~others.

\begin{figure}[t]
\centering
\includegraphics[width=0.45\linewidth,height=0.40\linewidth]{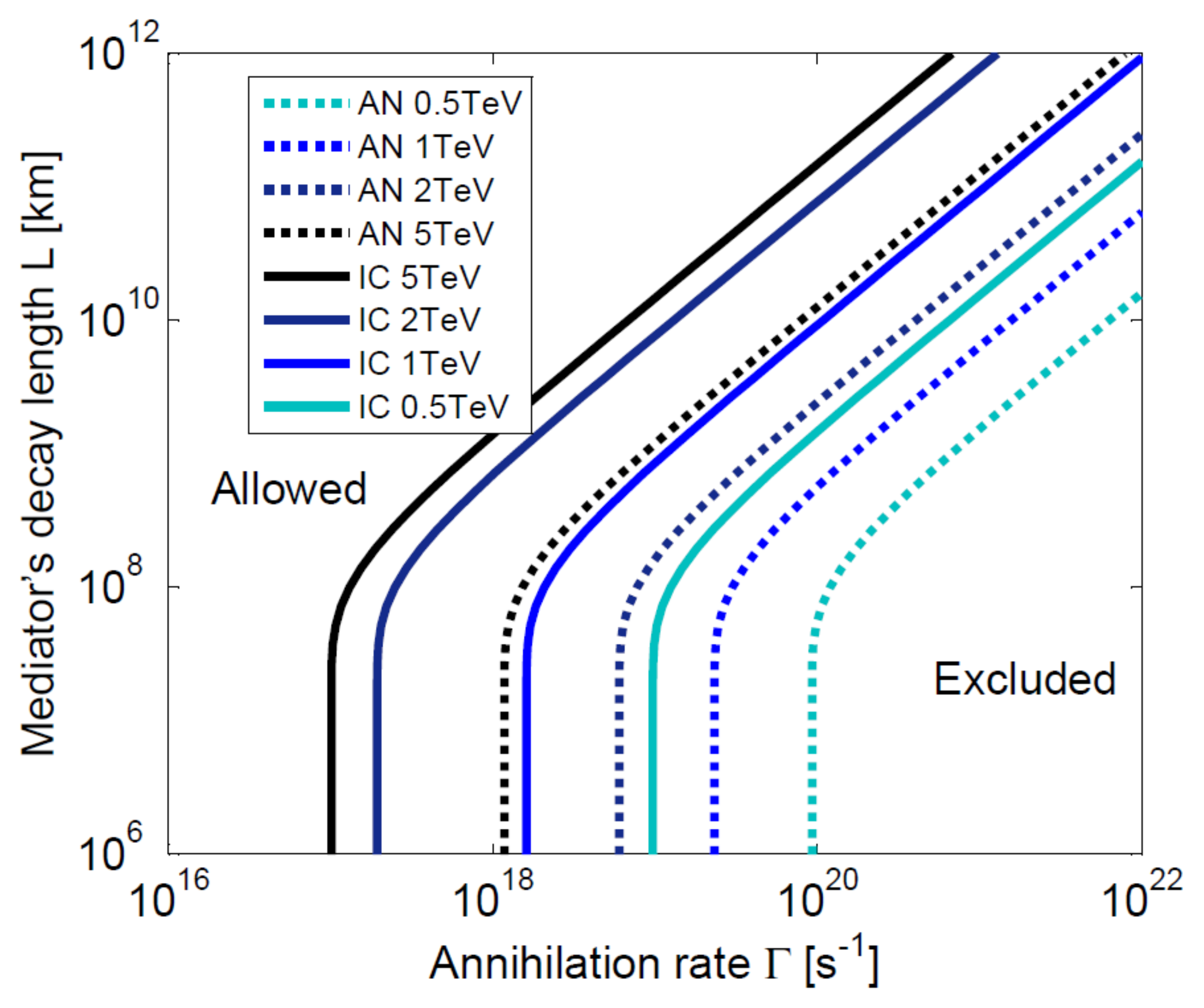} 
\includegraphics[width=0.45\linewidth,height=0.40\linewidth]{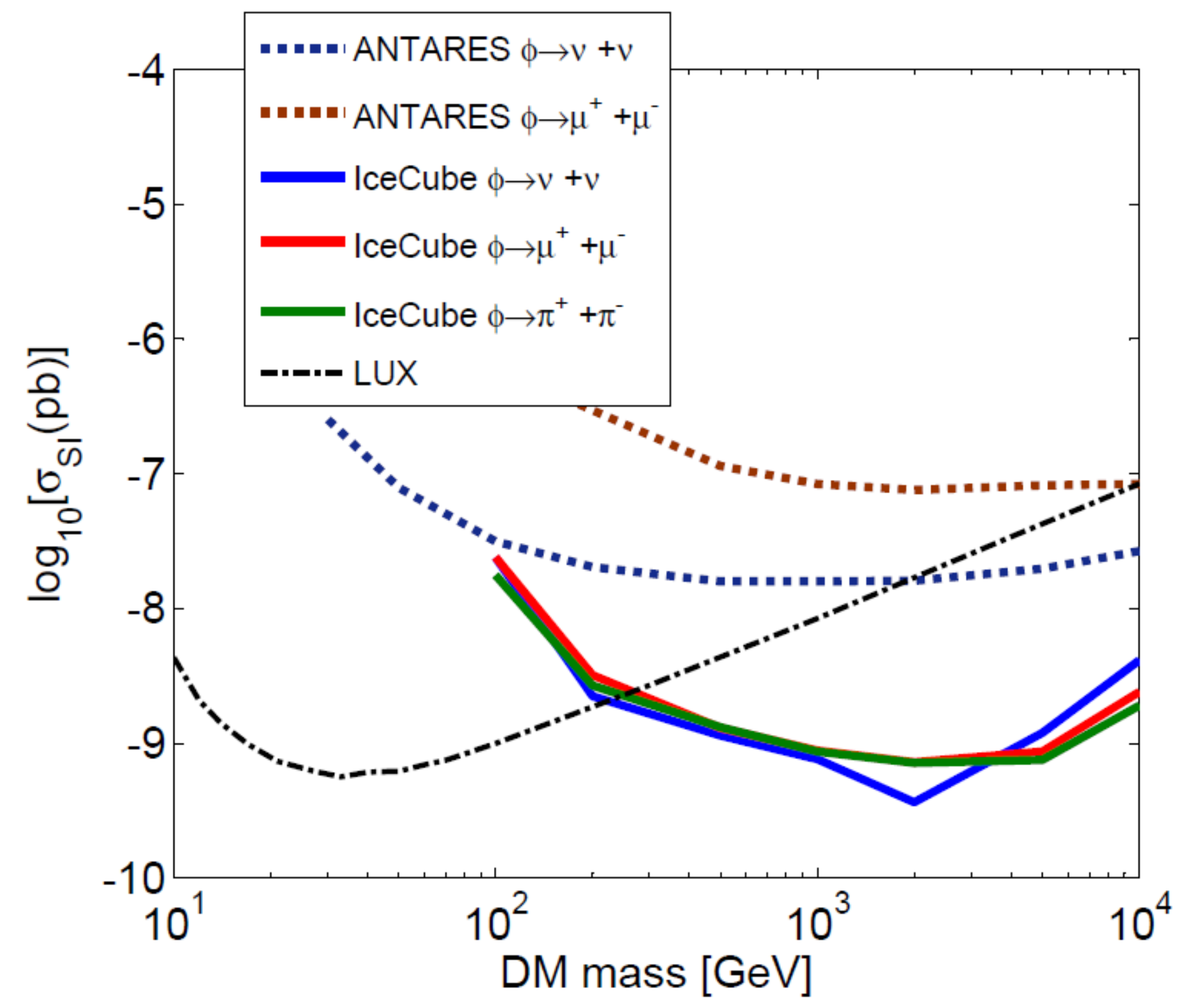} 
\caption{ (\textbf{Left}) The 90\% confidence level limits on the mediator decay length as a~function of the the annihilation rate in the Sun for a~several dark matter masses.
  The mediator is assumed to decay 100\% to neutrinos. Regions to the right of the curves are disfavoured. See the reference below for other decay modes.
  (\textbf{Right}) The 90\% confidence level limits on the spin-independent dark matter-nucleon cross section as a~function of dark matter mass for different mediator decay channels,
  assuming a~decay length of $2.8 \times 10^{7}$~km. Figures from~\cite{Ardid:2017lry}. Copyright IOP Publishing Ltd. and Sissa Medialab.  Reproduced by permission of IOP Publishing.  All rights reserved. }
\label{fig:DM_secluded}
\end{figure} 

There are some theoretical scenarios that predict a~flux of gamma-rays or cosmic-rays due to dark matter annihilations in the Sun. If~the dark matter ``world'' is coupled to our
usual world through long lived mediators that can escape the Sun interior and decay into Standard Model particles in-flight to the Earth, then other techniques can also search
for dark matter accumulated in the Sun. In~such models of {\it secluded} dark matter~\cite{Pospelov:2007mp},
an excess of gamma-rays or cosmic-rays from the direction of the Sun could also be interpreted as a~signature of dark matter annihilations inside it~\cite{Ajello:2011dq,Profumo:2017obk,Cuoco:2019mlb,Leane:2017vag}.
Secluded dark matter models also present advantages for neutrino searches. Neutrinos loose energy on their way out of the dense solar interior, so the flux that would reach
the Earth from annihilations in the solar core would be mainly low-energy (up to a~few 100~GeV) independent of the mass of the dark matter annihilating inside the Sun.
The correlation between dark matter mass and highest possible neutrino energy is therefore lost. However, since the decay of the mediator in secluded models can occur outside the Sun,
the~neutrinos will reach the Earth without loosing energy. This is advantageous for neutrino telescopes, which usually have better sensitivities and pointing for higher neutrino
energies, $\mathcal{O}$(100~GeV) and above.  These models add, of~course, additional free parameters: The mediator mass, lifetime and decay branching ratios to different final
states. Experimental results are therefore usually provided under different assumptions for the mediator mass and lifetime, for~a given annihilation
channel~\cite{Ardid:2017lry,Adrian-Martinez:2016ujo}. Figure~\ref{fig:DM_secluded} shows an~example of the results that can be attained within secluded dark matter models.
The figure shows limits on the mediator decay length as a~function of the the annihilation rate in the Sun (left plot) and limits on the spin-independent dark matter-nucleon
cross section as a~function of dark matter (right plot) obtained with public data of IceCube and ANTARES~\cite{Ardid:2017lry}. Note that neutrino telescopes,
IceCube in particular, provide the best limits on the spin-independent dark matter-nucleon cross section for dark matter masses above a~few 100~GeV for the model described
in the reference, improving even over direct search experiments like LUX, shown in the figure as the black dot-dashed~line.

\begin{figure}[t]
\centering
\includegraphics[width=0.70\linewidth]{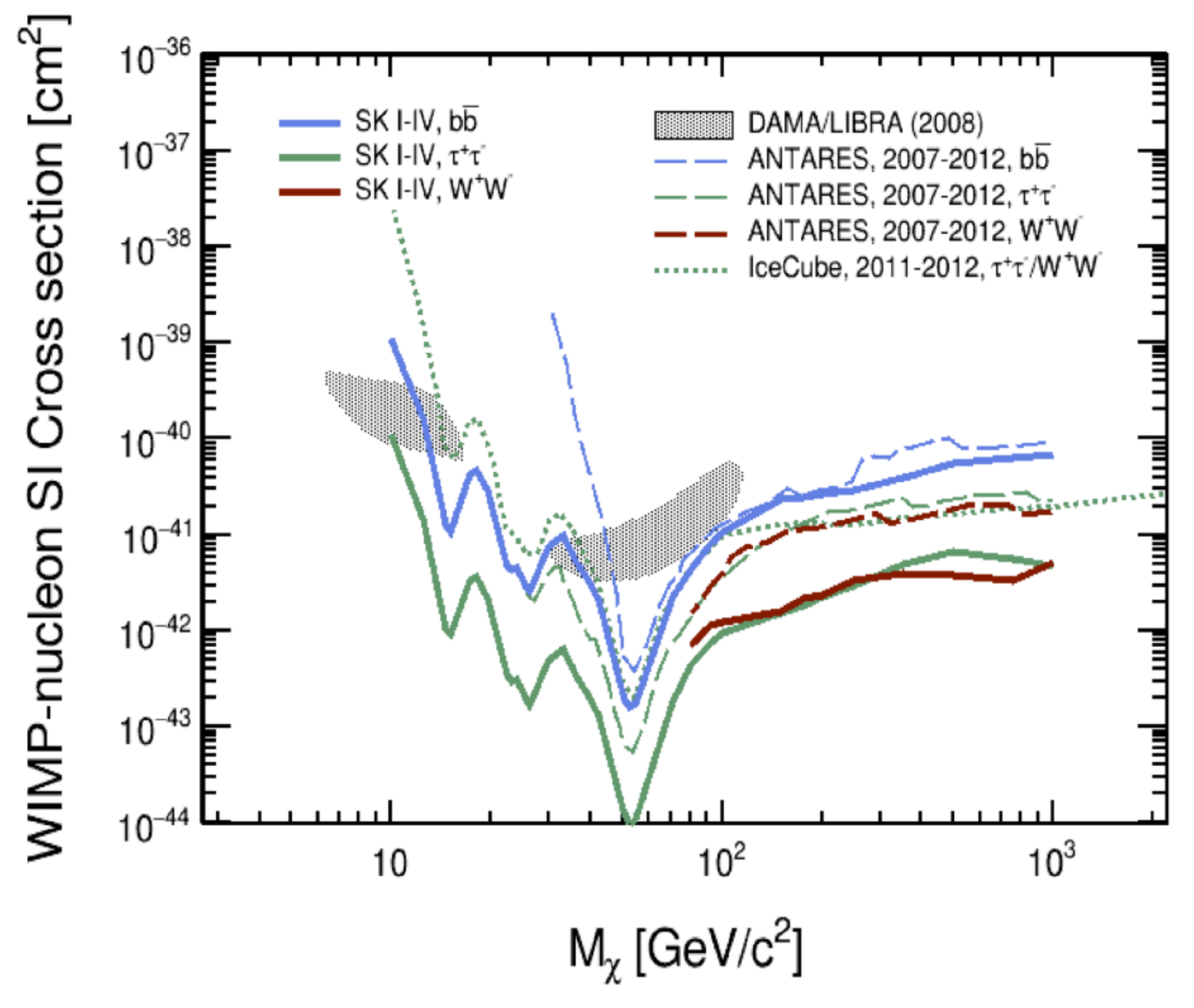} 
\caption{The 90\% confidence level upper limit on the spin-independent dark matter-nucleon cross section from annihilations in the Earth core for different annihilation channels. The~  shaded areas represent the best fit if one interprets the results from DAMA/LIBRA as indeed originating from dark matter interactions. Figure  from~\cite{Mijakowski:2020qer}.
  See references therein for details mentioned in the inset caption.}
\label{fig:DM_earth}
\end{figure}

The Earth presents somewhat a~different case, since the processes of capture of dark mater and annihilation in its core can not be assumed to have reached equilibrium.
Additionally, the~most abundant isotopes of the main components of the Earth inner core, mantle and crust, $^{56}$Fe, $^{28}$Si and $^{16}$O~\cite{Herndon:1980aaa},
are spin-0 nuclei, so capture in the Earth is a~good candidate to probe the $\sigma_{\rm SI}$ component of the dark matter-nucleon cross section. From~the experimental side,
the mass of these isotopes produces mass resonances in the capture rate, so neutrino searches for dark matter annihilations from the centre of the Earth are going to be
more sensitive to dark matter masses similar to the masses of those isotopes than to other masses. This is reflected in the shape of the limits obtained by IceCube, ANTARES
and Super-Kamiokande, shown in Figure~\ref{fig:DM_earth}. The~figure shows current limits on the spin-independent dark matter-nucleon cross section as a~function of
dark matter mass~\cite{Mijakowski:2020qer}.

\begin{figure}[t]
\centering
\includegraphics[width=0.75\linewidth]{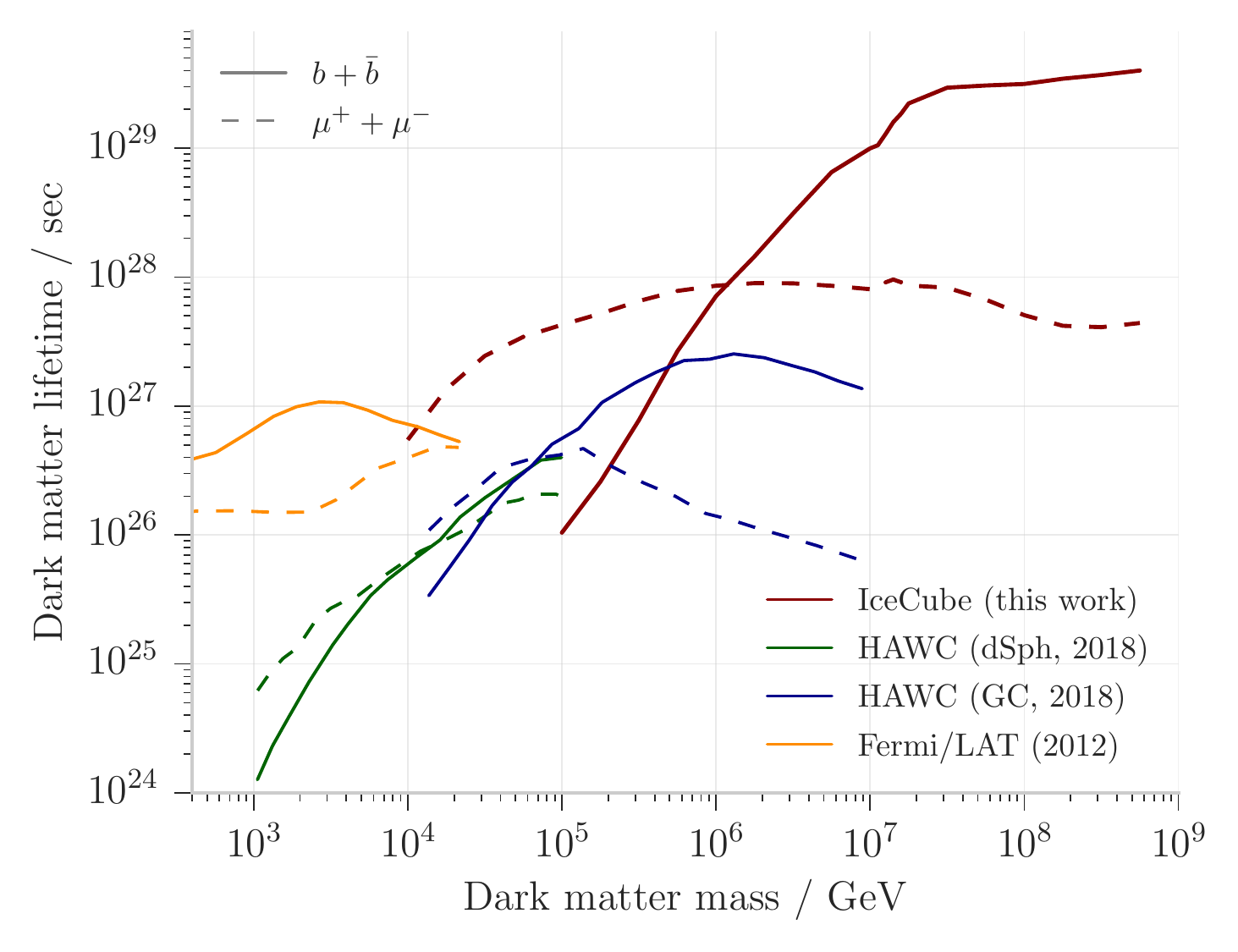} 
\caption{ Comparison of IceCube lower lifetime limits on heavy dark matter with results obtained by HAWC (Dwarf Spheroidal Galaxies and 
  Galactic Halo/Centre) and Fermi/LAT available at the time. Figure reproduced with permission from~\cite{Aartsen:2018mxl}. References to Fermi and HAWC data therein.}
\label{fig:DM_lifetime}
\end{figure}

However, perhaps of special interest is the recent discovery of a~high-energy cosmic neutrino flux by IceCube~\cite{Aartsen:2013jdh,Aartsen:2014gkd,Kopper:2017zzm}, since it
opens new possibilities for dark matter scenarios, specifically for annihilations or decays of very heavy dark matter, $\mathcal{O}$(PeV), that is impossible to probe in accelerators.
 These are called top-down scenarios, where high energy neutrinos are produced from the decay of heavy relic particles, as~opposed to bottom-up scenarios where cosmic neutrinos
 are expected to be produced from the decay of accelerated protons in a~source or its surroundings. Additionally,~extremely~heavy (and extremely light) dark matter scenarios are becoming
 of interest since classic WIMP models have been disfavoured by accelerator results and the persistent negative results of direct searches. 
Indeed even the first two PeV cosmic neutrinos detected by IceCube were promptly interpreted within the framework of PeV mass decaying dark matter candidates (R-parity violating gravitinos,
extra dimension fermions or hidden sector gauge bosons)~\cite{Feldstein:2013kka}. Decay of such heavy particles into neutrinos would produce a~monochromatic line, compatible with
the IceCube events at the time. With~currently an~excess of neutrinos well beyond 7$\sigma$ over the atmospheric neutrino background above 60 TeV, and~no sources identified except for
the blazar TXS 0506+056~\cite{IceCube:2018dnn,IceCube:2018cha}, the~interpretation of the IceCube excess as a~signature of dark matter has spurred a~lot of activity (e.g.,~\cite{Esmaili:2013gha,Bhattacharya:2019ucd,Boucenna:2015tra,Choi:2019ixb,Arguelles:2017atb,Kachelriess:2018rty,Chianese:2017nwe,Bhattacharya:2016tma,Dev:2016qbd,Murase:2015gea,Rott:2014kfa,Zavala:2014dla} among others). The~most general approach is to assume a~two-component flux where a~fraction of the IceCube neutrino flux is
assumed to originate from unidentified astrophysical sources, and~a fraction of the flux comming from heavy dark matter decays. The~low statistics of the IceCube data above 1~PeV do not allow for the moment to establish if the flux presents a~cut-off or not. A~sharp cut-off in the neutrino energy spectrum would favour the dark matter interpretation. Note that dark matter decays would produce also a~flux of high energy photons that can be recycled to lower energies through pair production and inverse Compton scattering on ambient radiation. So any model that explains the IceCube cosmic flux from heavy dark matter decay must be consistent with measurements of the gamma-ray sky, as~analyses like~\cite{Murase:2012xs,Cohen:2016uyg} show. Or, vice~versa, results from gamma-ray measurements can
constrain the decaying dark matter interpretation of neutrino and cosmic ray results within some dark matter models. 
Current experimental results on the lifetime of heavy dark matter from IceCube~\cite{Aartsen:2018mxl}, compared with limits from HAWC and Fermi-LAT, are shown in Figure~\ref{fig:DM_lifetime}.
The IceCube result assumes a~two-component fit, with~the astrophysical component following a~single power law. Lifetimes below 
$10^{27}$--$10^{29}$~s are disfavoured at 90\% CL, depending on the candidate~mass. 

\begin{figure}[t]
\centering
\includegraphics[width=0.75\linewidth]{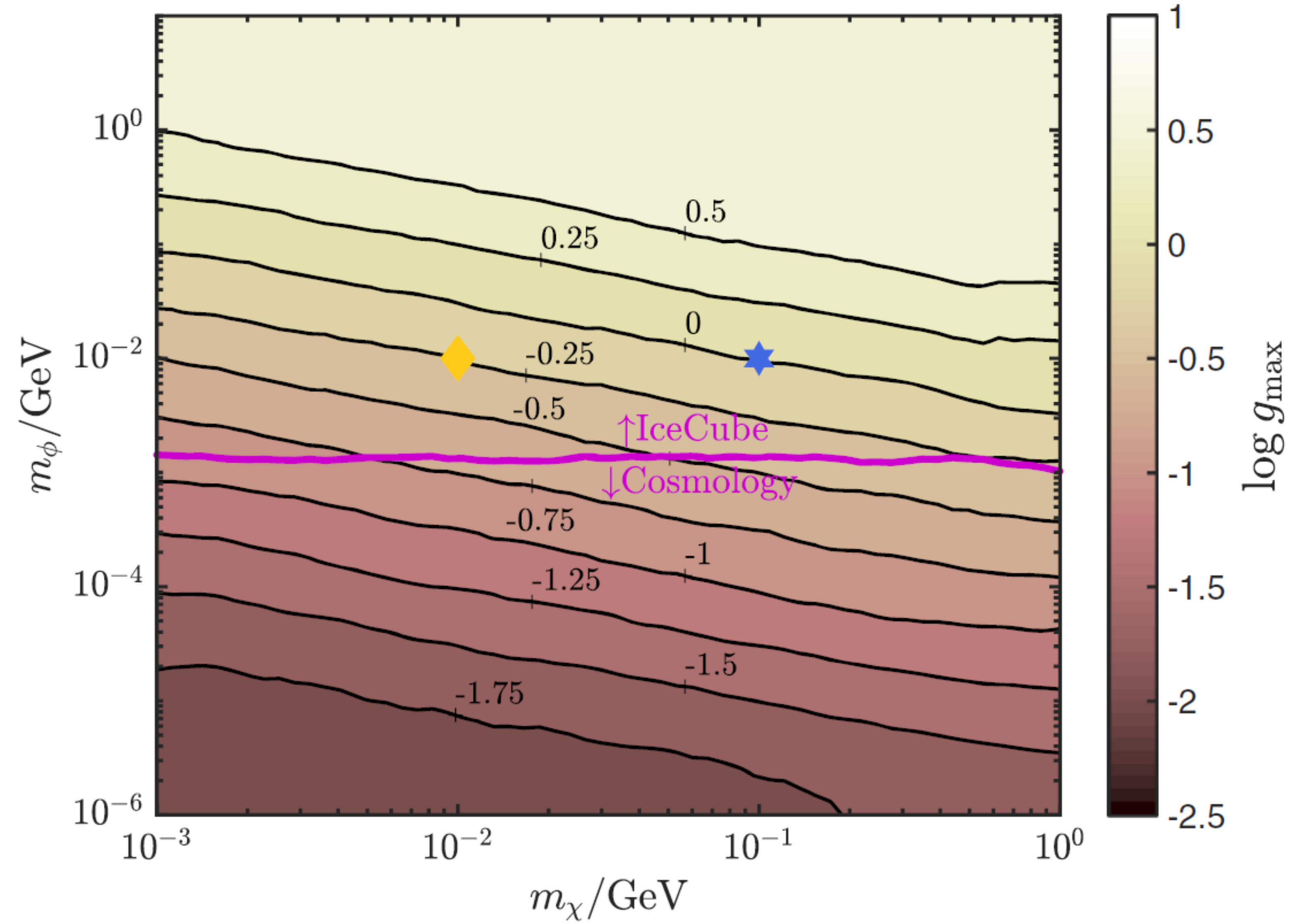} 
\caption{Maximum allowed value (color code) of the (log of the) coupling $g$ between neutrinos and dark matter, as~a function of dark matter mass, $m_{\chi}$, when the interaction is
  mediated by a~new vector boson of mass $m_{\phi}$. The~horizontal magenta line shows the division between the parameter space regions where a~neutrino telescope
  like IceCube can set stronger limits than analyses based on the power spectrum of the cosmic microwave background, or~vice~versa. Reprinted with permission from~\cite{Arguelles:2017atb}. Copyright 2020 by the American Physical Society.}
\label{fig:m_phi_vs_m_chi}
\end{figure}

The existence of a~high-energy cosmic neutrino flux can be used to probe conventional (i.e.,~not~extremely heavy, and~even very low mass) dark matter in another, neat, way. Astrophysical~neutrinos
reach the Earth from far away, traversing dark halos of galaxies in their way, including the Milky Way. Interactions of these neutrinos with the dark matter could reduce the
expected neutrino flux from specific directions, providing a~unique way of probing the neutrino coupling to the dark sector~\cite{Choi:2019ixb,Arguelles:2017atb}. 
While strongly model-dependent, these approaches can shed light on specific aspects of neutrino-dark matter interactions. 
An example of the type of results that can be obtained with these analyses is illustrated in Figure~\ref{fig:m_phi_vs_m_chi}. The~plot shows the result of a
search for a~depletion of cosmic neutrinos from the direction of the Galactic Centre due to neutrino-dark matter interactions. The~results are cast in terms of the mass 
of a~vector-like mediator of the interaction and the dark matter mass, as~well as the logarithm of the coupling strength of the interaction (color~code). Note that this method
is sensitive to very low mass dark matter candidates, on~the MeV-GeV range, and~therefore complementary to other analyses based on the cosmic neutrino flux. The~plot also shows that
neutrino telescopes have access to an~area of the parameter space, typically high mediator masses and high coupling constants, where limits from cosmological arguments
based on analyses of the cosmic microwave background do not~reach. 

\subsection{Cosmic-Rays}

Cosmic-rays (protons, electrons or light nuclei and their
antiparticles) can also be produced in the hadronization of $q\bar{q}$ or $W^+W^-$ pairs from dark matter annihilations or decays. However, using cosmic-rays as probes of dark matter poses challenges as
well. In~order to have access to particle identification, charge separation and energy measurement one must deploy real particle physics detectors in space, ideally with calorimetry and
a magnetic field for charged-particle tracking capabilities, a~technical challenge by itself. However, AMS~\cite{Lubelsmeyer:2011zz}, DAMPE~\cite{Fusco:2019lhc} or CALET~\cite{Torii:2019wxn} have demonstrated the feasibility of this type of detectors and produced impressive physics
results. The~limitations of using cosmic-rays as dark matter probes arise, though, from~fundamental physics. Cosmic-rays are charged particles (neutrons being too short lived to be of 
relevance) and propagation through the Galaxy can randomize their arrival directions and prevent accurate pointing to a~given object. Additionally, cosmic-rays can also lose energy through
inverse Compton scattering and synchrotron radiation during propagation, so their energy spectrum at Earth might differ from that where dark matter annihilations take place.
However, the detection of antimatter (antiprotons, positrons or light antinuclei) still provides a~good handle in dark matter searches since antimatter from astrophysical sources is
relatively rare at the Earth. Due~to its high  probability of annihilation with normal matter, any antimatter source needs to be relatively close to the Earth. However, this covers the
Galactic Centre and Halo, so there are plenty of predictions of what a~detection of an~excess of antiprotons or positrons would mean in terms of dark matter. However, even with antimatter
searches one has to deal with background from close-by sources, like~pulsars, which can produce an~(anti)cosmic-ray flux that can be indiscernible from a~dark matter signal.
Indeed the positron flux (compared to the electron flux) has been the source of controversy since results from the PAMELA satellite~\cite{Adriani:2008zr,Adriani:2013uda,Adriani:2017bfx} and the 
ATIC~\cite{Chang:2008aa} balloon experiment showed an~excess of positrons at energies above $\sim$10~GeV over the expectation from standard galactic primary cosmic-ray 
propagation and interactions, including the production of secondary electrons and positrons~\cite{Moskalenko:1997gh,Delahaye:2010ji}.
AMS02~\cite{Accardo:2014lma,Aguilar:2019owu}, CALET~\cite{Adriani:2017efm} and DAMPE~\cite{Ambrosi:2017wek} confirmed the excess and showed that
it extends to the TeV region, but~with a~cutoff at around 1~TeV. The~shape and normalization of the positron excess have given rise to a~wealth of models explaining the excess from dark matter
annihilations. 
Some of these solutions are quite contrived since, for~example, the~antiproton flux does not show such a~pronounced deviation from the expected background (there remains to be seen if there is
an antiproton excess at all, see below) and therefore an~explanation of the positron excess based on dark matter annihilations should prevent the same dark matter to produce a~significant excess
of antiprotons. These have been called leptophilic models, e.g.,~\cite{ArkaniHamed:2008qn,Haba:2010ag,Cohen:2009fz,Ibarra:2009bm,Fox:2008kb} 
and they are an~example of the challenges posed by trying to explain spectral features by invoking multi-component contributions to the spectrum. A~fit will always get better the more free
parameters are added to the function to fit, without~this reflecting on the physical relevance of the parameters. 
For anyone willing to wield Occam's razor, there is a~simpler, standard astrophysical explanation to a~positron excess in the cosmic-ray flux: It can indeed be produced by localized
pulsar wind nebulae close enough to the Earth~\cite{Hooper:2008kg,Profumo:2008ms,Malyshev:2009tw,Blasi:2010de,Yin:2013vaa,Bartels:2015aea}. In~a pulsar wind nebula, positrons are created
by electromagnetic cascading from seed electrons that have been ejected from the rapidly spinning neutron star. This electron and positron ``wind'' eventually creates a~shock when interacting
with the older ejecta of the original supernova remnant. A~fraction of the positrons might escape this complex system into the interstellar medium and become the galactic positron flux.
Pulsars are, thus, antimatter sources in the Galaxy. There are known, close enough pulsars to the Earth that can provide the needed positron flux to explain the measured excess over the
vanilla, point source free, background. Additionally, a~pure dark matter explanation of the positron excess is also 
strongly disfavoured by the fact that it should have produced a~signal in gamma-rays.  The~lack of a~signal in Fermi-LAT searches for dark matter in dwarf spheroid galaxies have set a~limit on the
dark matter annihilation cross section into channels that would produce an~$e^+e^-$ flux which is incompatible with the measured excess~\cite{Lopez:2015uma,Chan:2015gia}. Again, very few and very
contrived models can escape the strong Fermi-LAT constrains and still be compatible with the measurements of PAMELA, ATIC, AMS, CALET and~DAMPE.  

\begin{figure}[t]
\centering
\includegraphics[width=0.45\linewidth,height=0.40\linewidth]{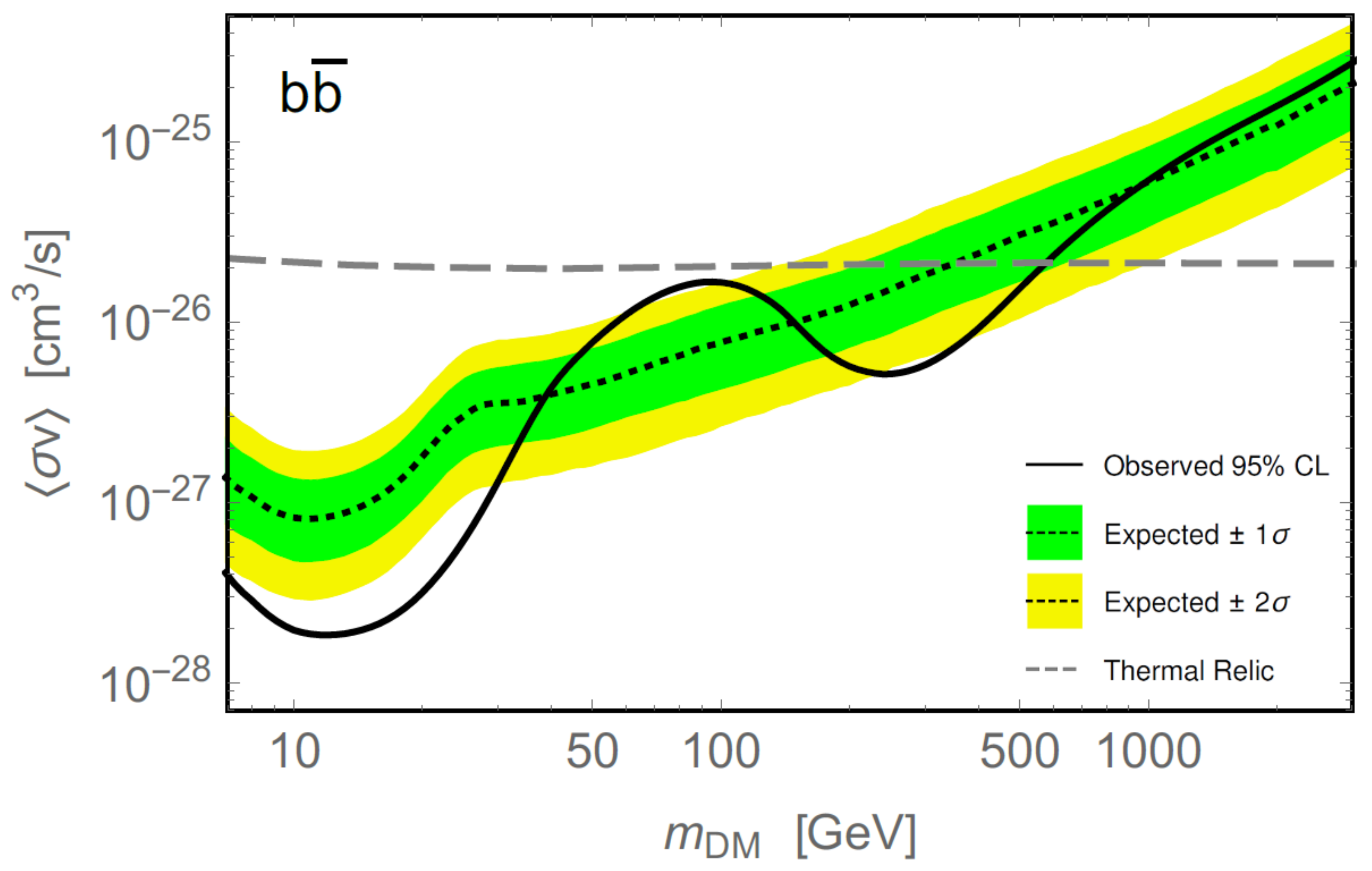} 
\includegraphics[width=0.45\linewidth,height=0.40\linewidth]{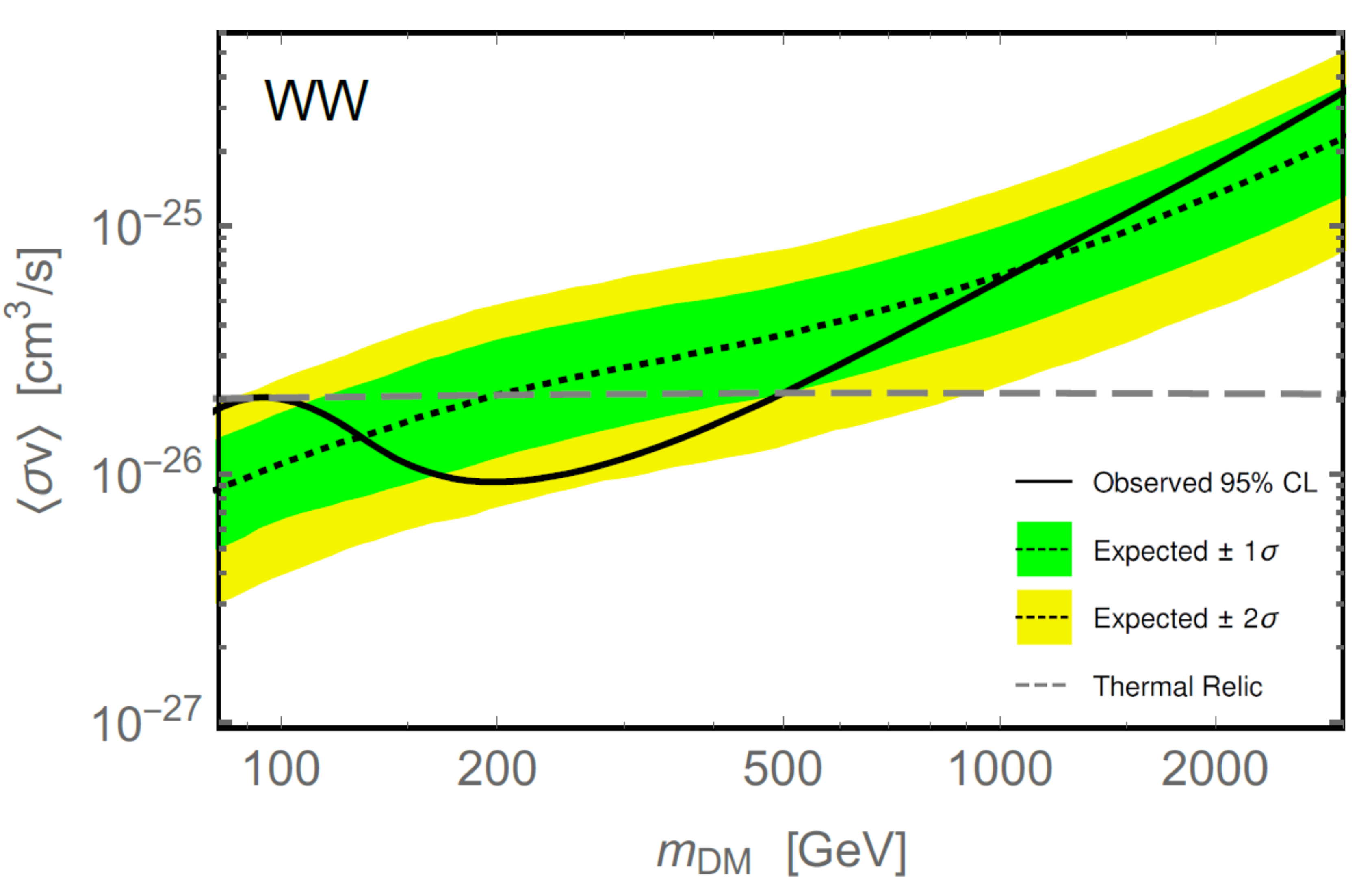} 
\caption{The 95\% confidence level limit on the dark matter annihilation cross section into $b\bar{b}$ (\textbf{Left}) and $W^+W^-$ (\textbf{Right}) as a~function of dark matter
  mass derived from the antiproton and B/C data of AMS-02. Figures from~\cite{Reinert:2017aga}. Copyright IOP Publishing Ltd. and Sissa Medialab.  Reproduced by permission of IOP Publishing.
  All rights reserved. }
\label{fig:limits_AMS}
\end{figure}

We mentioned above antiprotons, and~the story with them as a~dark matter signal is not so simple as it might have been suggested. 
Antiprotons are produced in interactions of protons and light nuclei with the interstellar medium. So, given the cosmic-ray flux, the~distribution of matter in the Galaxy, the~antiproton
production cross section and a~model for propagation, the~standard background of antiprotons can be, in~principle, calculated. Antiprotons from dark matter annihilations would be produced
as final state hadrons in annihilations to $q\bar{q}$ or $W^+W^-$ pairs, and~they would appear as an~
excess over the standard background~\cite{Bergstrom:1999jc,Donato:2003xg,Fornengo:2013xda}. Figure~\ref{fig:limits_AMS} shows the result of looking for dark matter annihilations into antiprotons with AMS-02. Limits on the velocity-weighted annihilation cross section as a~function of dark matter mass for annihilation into $b\bar{b}$ (left) and $W^+W^-$ (right) are shown, along with the 1$\sigma$ and 2$\sigma$ sensitivity~bands.

One current problem is that the uncertainties in the ingredients just mentioned above that enter into the calculation of the standard background  are larger than present 
measurements of the antiproton flux by AMS02. The~strongest culprit is the uncertainty on the antiproton production cross section in p-p and p-He collisions, which is just known to about 20\%,
slightly depending on energy, from~recent measurements of NA61 and LHCb~\cite{Korsmeier:2018gcy}. 
 Additionally, uncertainties in the propagation in the Galaxy also play a~determinant role. Antiprotons, as~charged
 particles, diffuse in the galactic magnetic field and a~complex transport equation, which must include diffusion, convection and reacceleration, with~uncertainties in each term, needs to be
 solved numerically to estimate the flux at the Earth.
 While there are recurrent claims of a~dark matter signal in cosmic-rays from the global fit industry~\cite{Cuoco:2019kuu,Cholis:2019ejx}, the~fact remains that, within~current uncertainties
 on the cosmic-ray propagation and on the elementary production cross sections, the~data remain  compatible with the expected background from standard processes. 
The simple fact that different authors performing what in principle is a~very similar analysis~\cite{Cholis:2019ejx,Reinert:2017aga} reach contradicting conclusions on the significance of
the antiproton excess (4.7$\sigma$ versus 2.2$\sigma$) points to the effect of the uncertainties involved in these type of calculations, mainly due to the cosmic-ray propagation in the
galaxy.

 A way out of this impasse might be the precise measurement of the antideuteron and antihelium fluxes, which should be correlated with the antiproton flux.
 A comprehensive analysis of the different antimatter cosmic-ray nuclei could help disentangle the standard or exotic origin of these fluxes~\mbox{\cite{Korsmeier:2017xzj,Reinert:2017aga}}.

\begin{figure}[t]
\centering
\includegraphics[width=0.45\linewidth,height=0.40\linewidth]{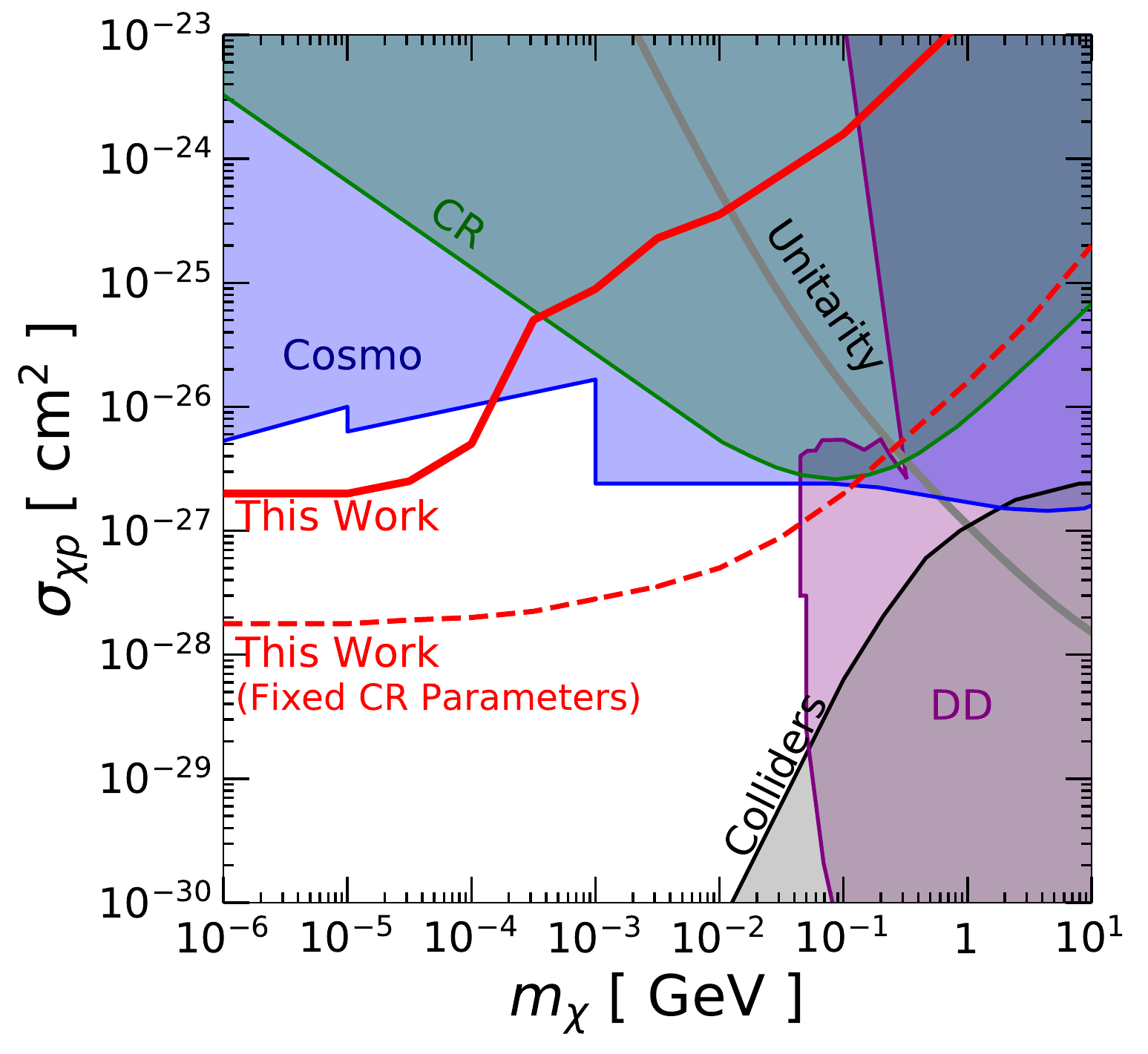} 
\includegraphics[width=0.45\linewidth,height=0.40\linewidth]{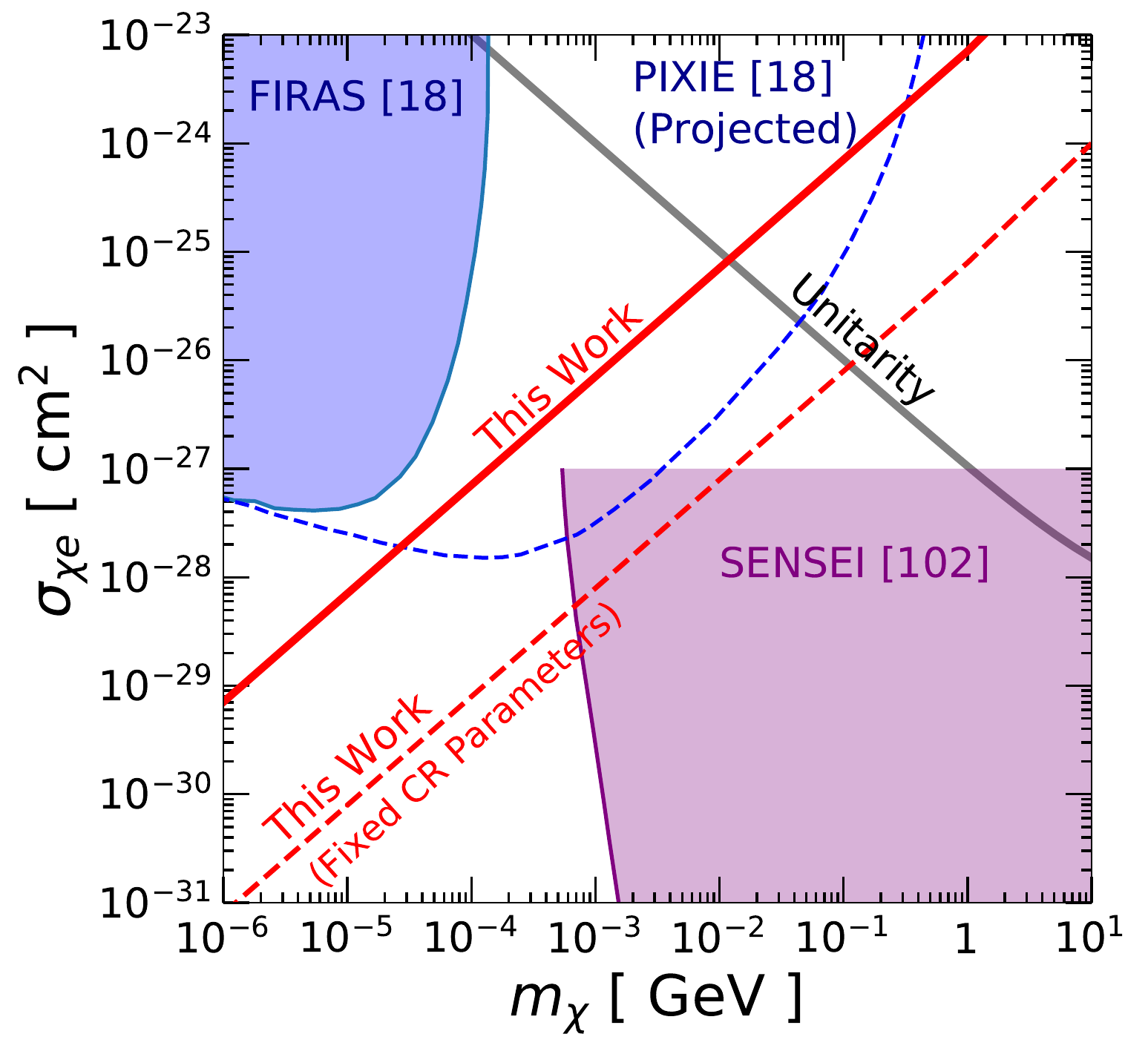} 
\caption{The 5$\sigma$ upper limit (red lines) on the dark matter elastic scattering cross section on protons (\textbf{left}) and electrons (\textbf{right}) obtained by comparing cosmic ray data from AMS-02 and CREAM with the expectation of cosmic ray
  propagation in the Milky Way including scattering on dark matter. Figures from~\cite{Cappiello:2018hsu}.}
\label{fig:sigma_invert}
\end{figure}

 A very different approach to use cosmic rays as tracers of dark matter is to look for deviations of the expected galactic flux of electrons and protons at the Earth due to interactions with
 dark matter during their propagation through the galaxy. These interactions will result in a~loss of energy of the cosmic rays and a~distorted energy spectrum at Earth in comparison with
 the scenario with no dark matter. This method has been used to  set limits on the {elastic} dark-matter proton and electron cross sections in what has been called ``inverse
 indirect detection''~\cite{Cappiello:2018hsu}. The~argument rests on a~similar idea than the one used with neutrinos in~\cite{Arguelles:2017atb} and illustrated in Figure~\ref{fig:m_phi_vs_m_chi},
 but using the galactic cosmic rays as ``beam'' and the dark matter particles in the galaxy as ``target''. The~method is specially sensitive to low dark matter candidates, well below a~GeV,
 since their number density can then be high enough for several scatterings to occur during the propagation of a~cosmic ray, leaving an~imprint in their spectrum. This approach complements direct detection experiments
 that are not currently sensitive to dark matter masses below $\mathcal{O}$(GeV) due to target threshold effects. On~the other hand, the~prediction on the expected distortion of the
 $p$ and $e$ spectra depends on the details of the propagation of these particles through the galaxy, which introduces certain degree of uncertainty. In~any case, the~analysis performed in~\cite{Cappiello:2018hsu}
 obtains very competitive limits on the elastic proton and electron cross section with dark matter for dark matter masses as low as a~keV. The~results are shown in Figure~\ref{fig:sigma_invert}
 for protons (left plot) and electrons (right plot). The~method can be extended to probe more complex interactions, like inelastic scattering or velocity dependent interactions, but~at
 the expense of introducing further model~dependencies.

\unskip 

\section{Constrains from Cosmology: CMB, BBN and the 21-cm Hydrogen~Line}

In this section we summarize the effect of dark matter annihilations and decays in observables related to the evolution of the early universe. This is an~
 independent way to infer the presence and characteristics of dark matter, be it on its imprint on the cosmic microwave background,
on the production of primordial light elements through Big-Bang Nucleosynthesis (BBN) or on the 21-cm power spectrum of Hydrogen (see~\cite{Clark:2019gug} for a~review and references therein). 
The  annihilation of dark matter present in the primordial soup injects additional energy in the form of photons, 
leptons or hadrons into the expanding universe. These will heat, ionize and excite the medium leaving imprints in the expected angular power spectrum that, assuming a~given dark matter candidate,
can be calculated and compared with observations. Turning the argument around: The consistency of the $\Lambda$CDM model with observations provides a~strong constrain on the characteristics of any
new dark matter particle species.  These methods have the advantage that they are free from astrophysical uncertainties like halo models of galaxies or dark matter distribution in clusters.
They also probe a~quite ample range of the history of the universe, from~the recombination era (z~=~1000) to the end of reionization (z$\sim$6). Incidentally, a~similar line of reasoning can be used
to set limits on dark matter annihilations or decays into final states including electromagnetic radiation at later stages in the evolution of the universe. A~direct comparison of the predicted
photon flux with the extragalactic background glow of the universe (infrared, ultraviolet or even optical) can be used to set constraints to the characteristics of several dark matter candidates~\cite{Overduin:2004sz}.
This method is not free from astrophysical uncertainties since it is applied to the older universe, when structures have been formed, and~the expected signal depends on how the dark matter,
as well as dust and gas, is distributed in galactic halos or the intergalactic medium. The~limits obtained are also not as strong as more recent limits mentioned earlier in this~review.

\begin{figure}[t]
\centering
\includegraphics[width=0.45\linewidth,height=0.45\linewidth]{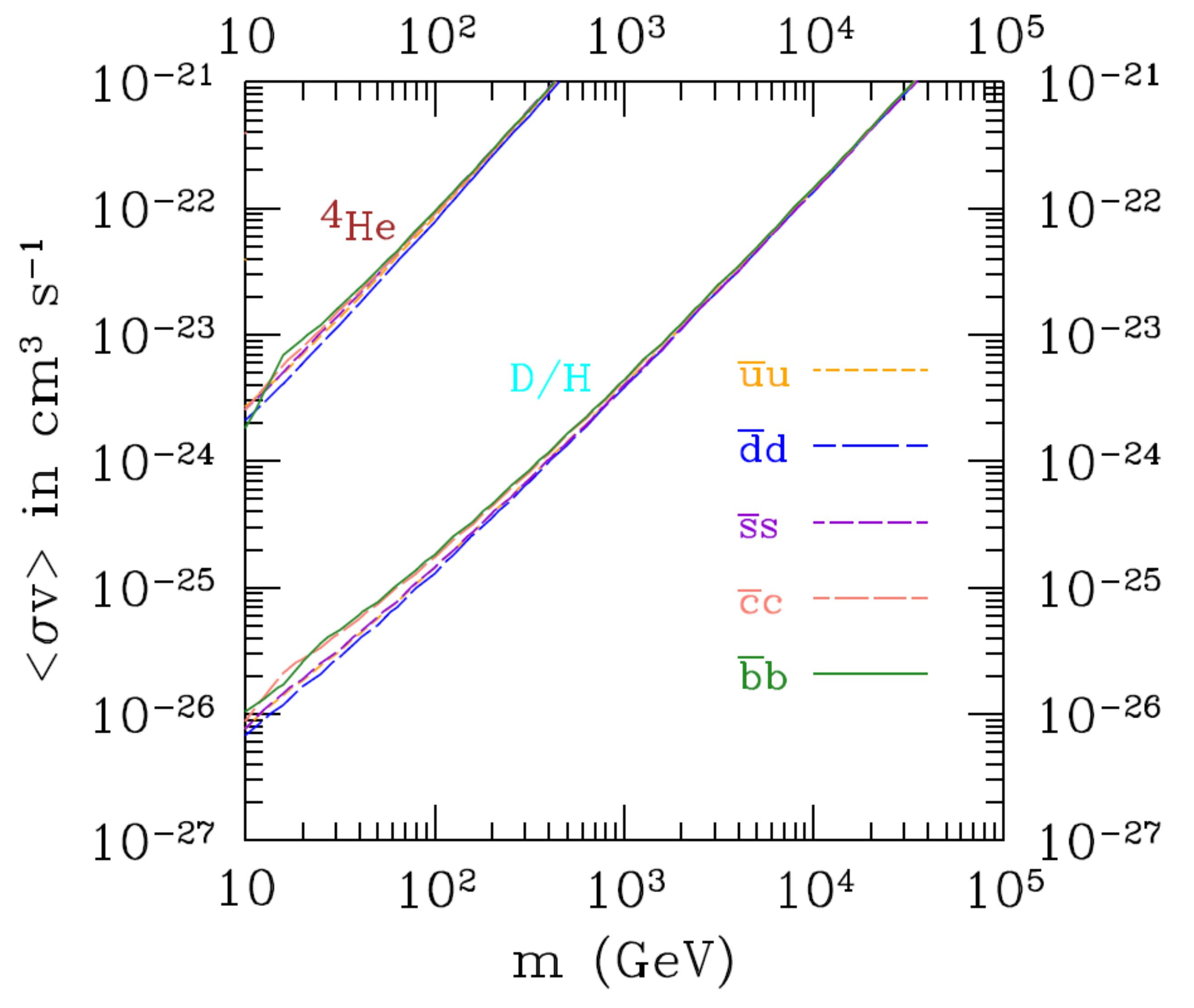} 
\includegraphics[width=0.45\linewidth]{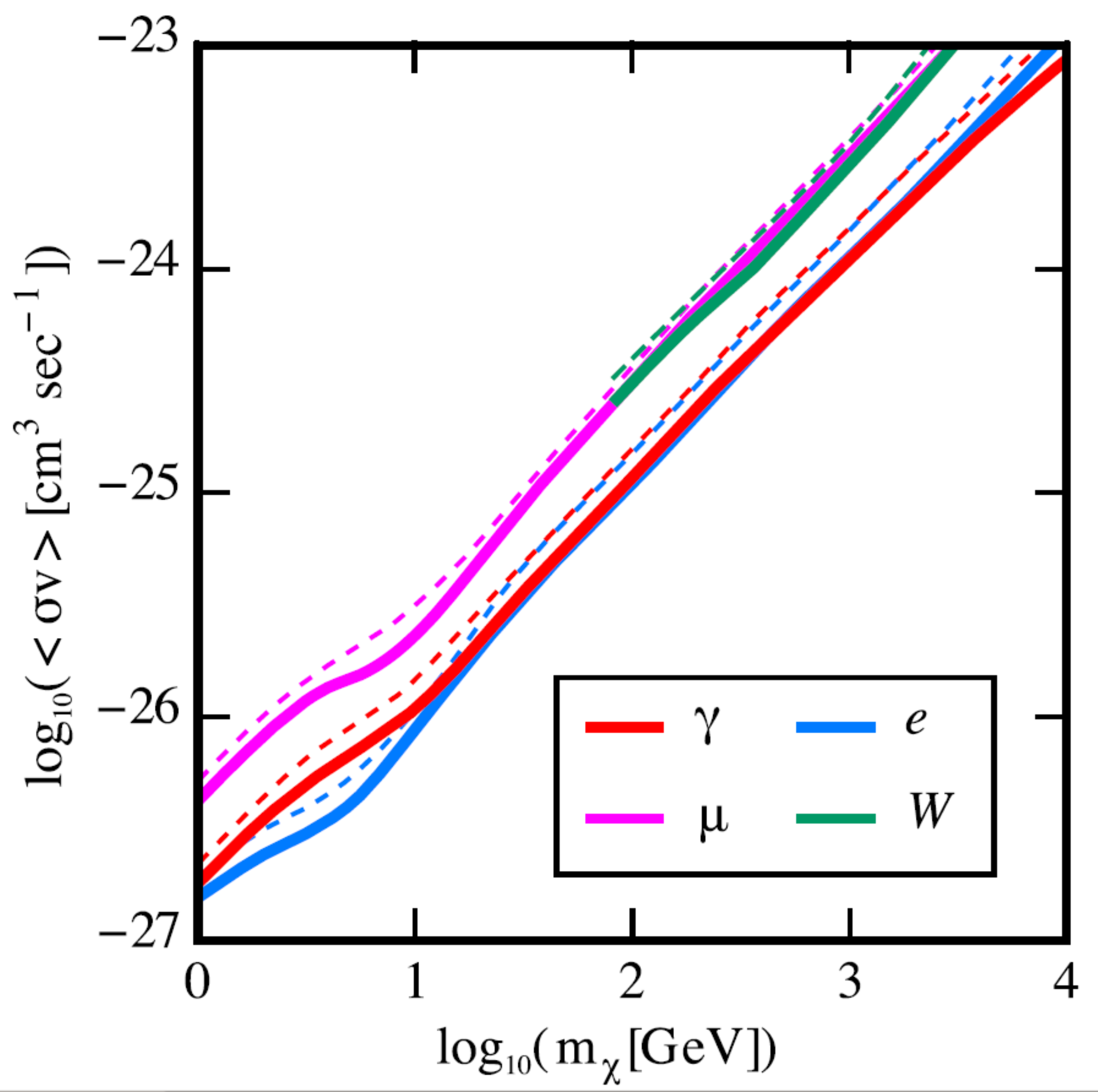} 
\caption{
(\textbf{Left}) The 95\% confidence level limits on the velocity-weighted dark matter annihilation cross section as a~function of dark matter mass for different annihilation channels,
  obtained from an~analysis on the primordial abundances of light elements from big-bang nucleosynthesis. Figure from~\cite{Kawasaki:2015yya}.
    (\textbf{Right}) The 95\% confidence level limits on the velocity-weighted dark matter annihilation cross section as a~function of dark matter mass for different annihilation channels,
obtained from an~analysis of CMB anisotropy data. Figure from~\cite{Kawasaki:2015peu}. 
}
\label{fig:CMB-BBN_limits}
\end{figure}

Following the evolution of the universe, from~higher to lower redshifts, the~first effect to consider is the imprint of the energy dumped into the primordial plasma by annihilations or decays
of dark matter, and~its effect on element nucleosynthesis.  Photons, electrons and positrons couple readily to the plasma, but~quarks or gluons will hadronize into strongly interacting particles, typically mesons and baryons (and their antiparticles). 
Short-lived particles (with lifetimes much shorter than the strong interaction rate at the temperature of the surroundings) will decay, reinforcing the energy dumping into the plasma and inducing
photo-dissociation of already formed light elements. However, stable particles, like~protons (or neutrons for all purposes) will become part of the expanding plasma, ready to participate in nucleosynthesis.  
Since light-element synthesis depends on the baryon-to-photon ratio, additional stable baryons in the plasma due to dark matter annihilations, added to the photo-dissociation mentioned above
and pion-exchange interactions that will drive proton-neutron conversion, 
can~change the rate of certain reactions and 
produce primordial element abundances incompatible with observations. Turning the argument around, the~observed abundance of primordial elements can serve to set limits on the
amount of dark matter annihilations during the nucleosynthesis era~\cite{Reno:1987qw,Kawasaki:2015yya,Jedamzik:2004ip,Hisano:2009rc}. The~main uncertainties of this procedure are the uncertainties in
measuring the primordial element abundances, a~difficult measurement due to the intrinsic contamination of non-primordial elements in any source considered. Measurements of some elements have reached
a precision of less than 1\% ($^{4}$He~or D), but~others like $^{7}$Li remain at about 20\%~\cite{Fields:2019pfx}. Even with these caveats, BBN arguments allow to set quite stringent limits on $\left < \sigma_{\rm A} v \right >$ as a~function of dark matter mass, as~it is shown in the right plot of Figure~\ref{fig:CMB-BBN_limits}.

 Changes in BBN processes will bear on the temperature and polarization power spectra of the CMB. The~CMB conveys detailed
information about the last scattering surface and, if~dark matter decayed or annihilated in the early universe, the~extra energy deposited in the plasma could delay recombination or
smooth out the primordial density fluctuations. Such an~effect would be visible in the CMB data today~\cite{Slatyer:2015jla,Slatyer:2015kla,Zhang:2006fr}. However, not 
all the energy released in the annihilation of two dark matter particles is absorbed into the surrounding medium and it is usual to express results from CMB analyses in terms
of an~efficiency factor, $f(z)$, which represents the fraction of the annihilation energy that goes into heating and ionizing the surrounding plasma. This factor depends both on redshift and
the specific dark matter model through the branching ratios to particles that couple to the plasma, typically~photons and electrons and positrons, but~it lies in the range 0.001--1~\cite{Ade:2015xua}. The~redshift dependence of $f(z)$ is less acute than its dependence on the dark matter model, since the universe is evolving fast, and~the effects of injecting
energy into it are relevant only during a~limited $z$ range, typically between 1000 and 600, so one can approximate $f(z)$ with an~effective constant value $f_{\rm eff}$ which practically
only depends on the properties of the dark matter model assumed. Following~\cite{Ade:2015xua}, the~energy released per unit volume from dark matter annihilations as the universe expands can
be written as
\begin{equation}
\frac{dE}{dt dV}(z)~=~2 g^2 \rho_{\rm crit}^2 c^2 \Omega_c^2 \left( 1+z \right)^6 f_{\rm eff} \frac{\left < \sigma_{\rm A} v \right >}{m_{\rm DM}}
\label{eq:CMB_injection}
\end{equation}
where $\rho_{\rm crit}$ the critical density of the universe today, $m_{\rm DM}$ is the mass of the dark matter particle, and~$\left < \sigma_{\rm A} v \right >$ is the thermally-averaged
annihilation cross-section times the dark matter velocity. While~results from CMB observations are sometimes given as a~function of $f_{\rm eff} \left < \sigma_{\rm A} v \right >$, detailed~calculations of the effect of annihilations assuming a~simple dark matter model can be performed in order to break the degeneracy and express the results as limits to the
$\left < \sigma_{\rm A} v \right >$ as a~function of the dark matter mass. This has the advantage of presenting results which are directly comparable with other indirect search techniques mentioned in
this review, while not being more model dependent. On~the contrary, rather less model dependent since, as~mentioned above, source modeling does not enter here aside from the
usual assumptions of $\Lambda$CDM cosmology. There is one caveat, though, when~comparing limits on  $\left < \sigma_{\rm A} v \right >$ from indirect experiments, probing dark matter
accumulated in galactic halos, with~limits on  $\left < \sigma_{\rm A} v \right >$ from CMB.
Note that the direct comparison works under the assumption of s-wave annihilations. For~higher order contributions or 
more complex models, with~velocity-dependent annihilation cross section for example, the~comparison of limits should be taken with care since they probe very different dark matter velocity regimes.
A thermalized 100~GeV particle in a~3000~K plasma (the temperature of the universe at recombination) has a~typical velocity of the order of 100~m/s, while dark matter particles in galactic 
halos have typical velocities of 200~km/s. 
The left plot of Figure~\ref{fig:CMB-BBN_limits} shows limits on $\left < \sigma_{\rm A} v \right >$ as a~function of dark matter mass, where a~detailed calculation of the effects in the CMB anisotropy induced by the
injection of photons, electrons, muons and W's from dark matter annihilations in the expanding plasma has been used to calculate $f_{\rm eff}$~\cite{Kawasaki:2015peu,Kanzaki:2009hf,Kanzaki:2008qb}. 

After recombination, the~universe enters a~phase where primordial atoms have been formed and the ambient radiation does not have enough energy to ionize them. This is called the
``Dark Ages'', and~it lasts until about z$\sim$200, when the first large structures start to form by gravitational growth from the primordial fluctuations. As~matter collapses into structures it heats
up, emitting ultraviolet radiation that can interact with the surrounding gas, specifically inducing a~strong absorption at the resonant frequency of the hyper-fine transition between the triplet and the singlet levels of the hydrogen ground state. These two levels are separated by 
$5.9 \times 10^{-6}$~eV, which corresponds to a~wavelength of 21.1~cm and a~frequency of 1420~MHz. Today, the~redshifted absorption signal appears typically in the 30--200~MHz range and opens the possibility to
use radio telescopes as instruments to study the early universe. The~21-cm line has been used to measure rotation curves of galaxies in a~complementary way to optical measurements. However, the primordial
21-cm sky follows the matter distribution at the reionization era and provides additional information to CMB cosmology~\cite{Pritchard:2011xb}. At~about z$\sim$6 reionization ends and the whole universe
remains ionized until the present epoch. However, up to redshifts about z$\sim$50, the~21-cm signal is expected to be negligible. The~presence of dark matter annihilations or decays can change this timeline,
providing a~continuous source of energetic electrons, positrons and gamma-rays that could start the reionization at an~earlier time and leave a~measurable effect in the 21-cm map~\cite{Evoli:2014pva,Belikov:2009qx,DAmico:2018sxd,Natarajan:2008pk,Natarajan:2009bm}. Ideally, one could even distinguish the time evolution of dark matter annihilations, which is different from decays. Since the annihilation rate is proportional to $N_{\rm DM}^2$, annihilations can dominate
at early times, when the cosmic expansion has not diluted the number density, but~they could reignite at a~later stage, when dark matter has collapsed into primordial halos, providing a~local enhancement
of the dark matter density. Since decays are just proportional to $N_{\rm DM}$, the~energy injection into the surrounding medium is uniform with time~\cite{Cumberbatch:2008rh,Chuzhoy:2007fg}. 
Given the wide energy spectrum of the annihilation/decay products of dark matter (up to the dark matter mass) and the resonance character of ionization, only a~small fraction of the injected energy
from dark matter can be used for ionization. This limits the dark matter models that can be probed by 21-cm line cosmology to low-mass, typically $\mathcal{O}$(MeV), candidates. Only a~tiny fraction of
the spectrum of annihilations of heavier candidates is effective for ionization, so the strength of the method decreases rapidly with dark matter mass~\cite{Ripamonti:2006gq,Ripamonti:2006gr,Mapelli:2006ej,Mapelli:2007kr,Lopez-Honorez:2016sur}.

As we showed for the case of the CMB, the~energy injected into the intergalactic medium due to dark matter annihilations can be related to the annihilation cross section and candidate mass. There~are several ways to express this, as~fraction of ionized Hydrogen as a~function of redshift, as~the temperature of the Hydrogen gas as function of redshift or, what I think gives a~more visual connection
to the dark matter characteristics, as~the rate of energy deposition from annihilations as a~function of redshift~\cite{Valdes:2007cu},
\begin{equation}
\frac{dE}{dt}(z)~=~\frac{1}{2} N^2_{\rm DM,0}  N_{\rm b,0} \left( 1+z \right)^3 f_{\rm abs}(z) \frac{\left < \sigma_{\rm A} v \right >}{m_{\rm DM}}
\label{eq:21cm_injection}
\end{equation}
where $N_{\rm DM,0}$ is the current number density of dark matter particles, $N_{\rm b,0}$ is the current baryon number density, $f_{\rm abs}$ is the fraction of the energy released by the dark matter
annihilations that goes into ionization and, as~in Equation~(\ref{eq:CMB_injection}), $m_{\rm DM}$ is the mass of the dark matter particle and $\left < \sigma_{\rm A} v \right >$ is the thermally-averaged
annihilation cross-section times the dark matter velocity. 
The calculation of the left hand side of Equation~(\ref{eq:21cm_injection}) is not easy since the injection of energy occurs in a~complex and expanding medium, but~
once a~model for the evolution of the intergalactic medium is in place, observations of the 21-cm primordial intensity in the sky (proportional to the injected energy from dark matter annihilations) can
be used to constrain $\left < \sigma_{\rm A} v \right >$ as a~function of $m_{\rm DM}$. 
The main foreground comes from the contribution from later processes, like radio emission from synchrotron processes in our galaxy or from other galaxies and clusters of galaxies, which has to
be carefully dealt~with. 

\begin{figure}[t]
\centering
\includegraphics[width=0.70\linewidth]{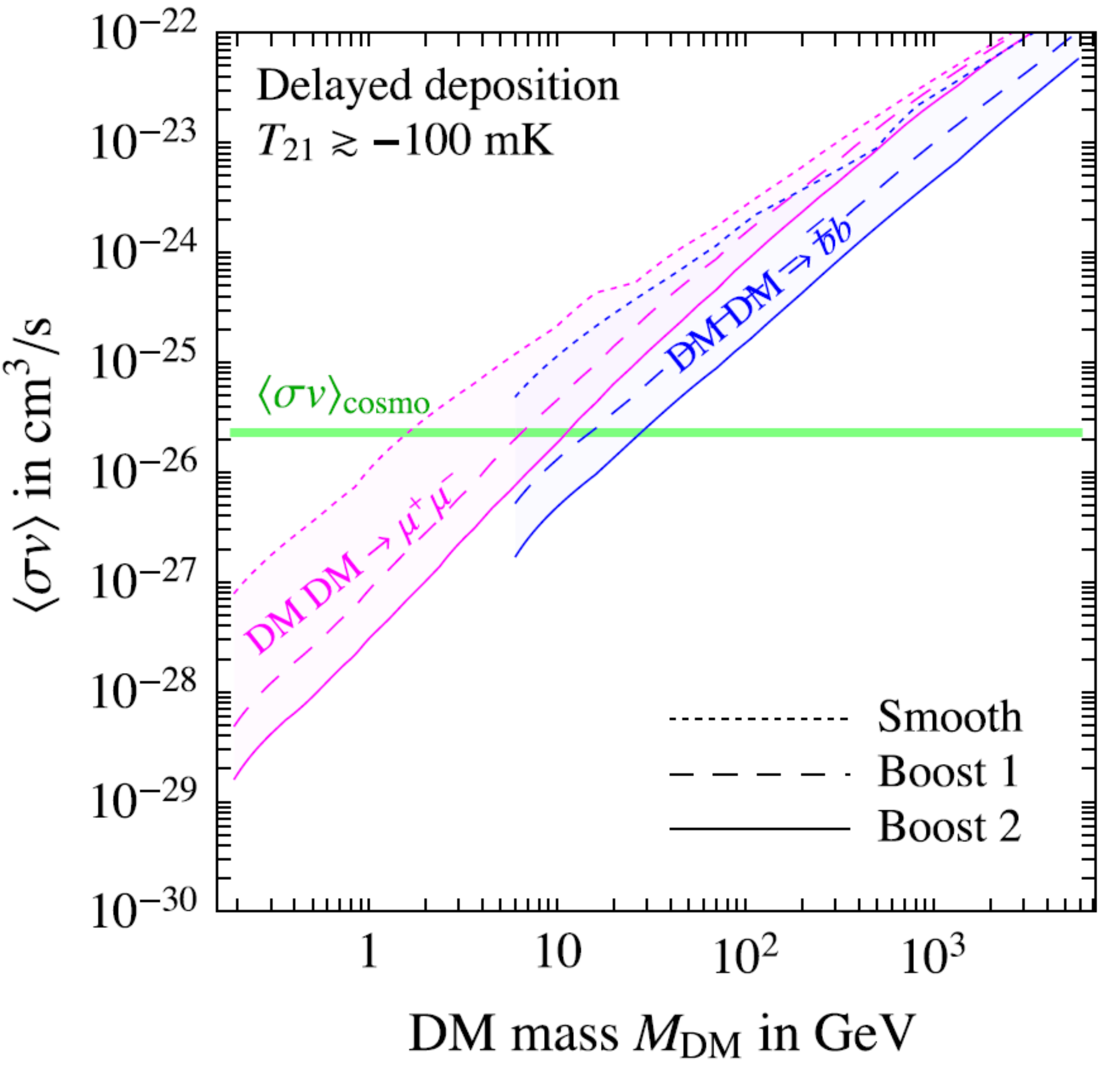} 
\caption{ Limits on the velocity-weighted dark matter annihilation cross section as a~function of dark matter mass from the absorption observed in the 21-cm spectrum at $z\sim 17$, for~  two examples of annihilation channels. Figure from~\cite{DAmico:2018sxd}. }
\label{fig:21cm_limits}
\end{figure}

There are several radio observatories in operation that have the sensitivity to set competitive limits on dark matter annihilations and, indeed, observations from the EDGES telescope~\cite{Bowman:2018yin}
have been used to set limits on dark matter characteristics. The~EDGES telescope measured the brightness temperature of the 21-cm line relative to the CMB, finding that the absorption profile is consistent with expectations. However, it also found that the relative amplitude of the absorption is more than a~factor of two larger than the predictions from standard cosmology at a~3.8$\sigma$ confidence level. Since we are talking of a~relative measurement of $T_{\rm 21cm}$ with respect to $T_{\rm CMB}$, the~discrepancy can be interpreted either as the primordial gas at z$\sim$20 being colder than expected, or~the background radiation being hotter than expected ($T_{\rm 21cm}$ and $T_{\rm CMB}$ denote, respectively, the~temperature of the ambient gas and of the CMB at the end of the Dark Ages). Standard explanations assuming increased radiation from the early stars and/or a~
slightly different decoupling redshift between matter and radiation can (more~or less) accommodate the EDGES measurement, since there are still large uncertainties on the early star formation history of
the universe. However, the result from EDGES can also be interpreted in terms of dark matter. Indeed, interactions of dark matter with surrounding baryons or the injection of energy due to dark matter decays or annihilations can change the thermal history of the hydrogen gas and accommodate the observation by EDGES~\cite{Panci:2019zuu,Liu:2018uzy,DAmico:2018sxd,Barkana:2018lgd}.
The contribution of dark matter annihilation products to heating the background radiation can lead to a~suppression of the observed absorption (even erasing it if the energy injection is too large), so
the observation of an~anomalous absorption level in the 21-cm line sky can be translated into bounds on dark matter annihilations or decays. On~the other hand, interactions 
between dark matter and surrounding  baryons can lower the ambient gas temperature, and~can explain the observed feature if the dark matter species is light (below a~few GeV) and its interaction cross
section lies above a~few $10^{-21}$~cm$^2$~\cite{Barkana:2018lgd}. Figure~\ref{fig:21cm_limits} shows an~example of the constraints obtained from the EDGES observation for two annihilation channels, and~assuming that the energy deposited from the annihilations is absorbed by the surrounding plasma with some delay~\cite{DAmico:2018sxd}. This needs to be modeled and adds a~source of uncertainty in the calculations but it is justified for some annihilation products that can penetrate the gas before fully interacting with it. In~other cases, like short-lived particles, a~``prompt'' energy deposition approximation is more accurate. See~\cite{DAmico:2018sxd} for additional limits obtained under the prompt energy deposition approximation. 
Note that the limits reach lower masses than other indirect searches, but~the caveat mentioned above about the different average velocities of dark matter today (as probed by annihilations in the halo of our Galaxy or nearby galaxies) and at higher redshifts (as probed by the 21-cm line) remains when trying to compare~results.

\begin{figure}[t]
\centering
\includegraphics[width=0.70\linewidth]{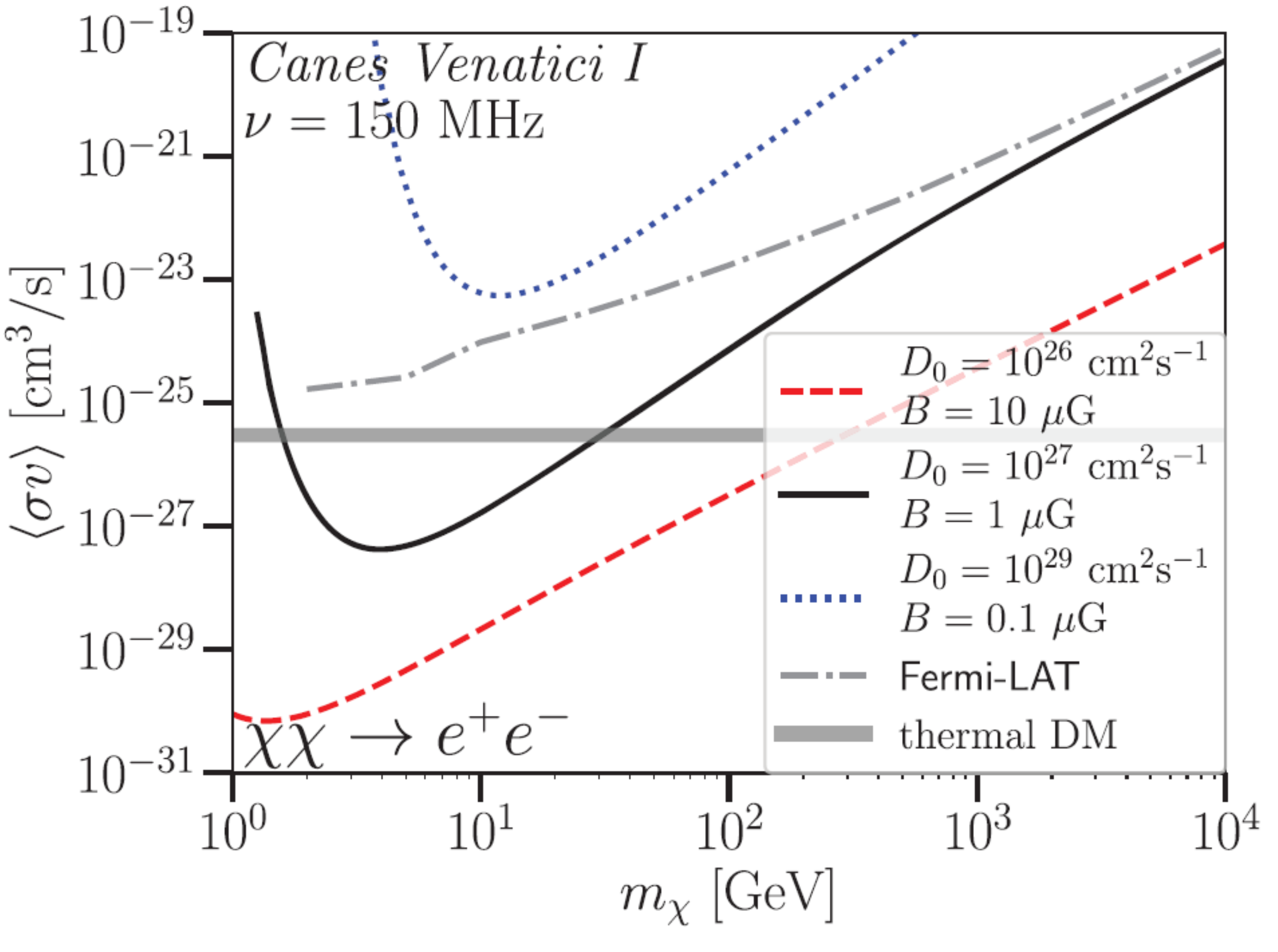} 
\caption{ Limits on the velocity-weighted dark matter annihilation cross section as a~function of dark matter mass from LOFAR, assuming annihilation into $e^+e^-$. Results are dependent on the galactic diffusion model used and assumed magnetic field strengths, and~results are shown for three different models (blue dotted, black full and red dashed lines). Figure from~\cite{Vollmann:2019boa}. See reference for details in the diffusion models~used. }
\label{fig:LOFAR}
\end{figure}

Constrains on the dark matter annihilation cross section can also be obtained by using radio telescopes to target specific sources, in~a similar way that it is done with gamma-ray, X-ray and neutrino telescopes. The~technique is based on looking for synchrotron radio emission from the charged products (mainly e$^-$ and e$^+$) of dark matter annihilations in the halo of a~suitable galaxy. The~author
in~\cite{Tyler:2002ux} used Draco as early as 2002 to set limits on the dark matter velocity averaged self annihilation cross section from VLA observations. More recently, the~LOFAR collaboration has  carried out a~competitive search for dark matter in the dwarf galaxy Canis Venatici~\cite{Vollmann:2019boa}. The~results of this technique depend on the diffusion of the charged particles once
emitted and on the profile of the local magnetic field, two quantities that are not well known, so the results are ineluctably model-dependent. However, with reasonable assumptions on the structure and
magnetic field strength from a~given source, this~method yields competitive results to gamma-ray, cosmic-ray or neutrino dark matter searches from dwarf galaxies as illustrated in Figure~\ref{fig:LOFAR}.

\unskip

\section{Discussion}

 Even that severe constraints from accelerator experiments have closed a~big chunk of the supersymmetric parameter space, which was
originally a~favourite to provide a~dark matter candidate with the right mass and interaction strength, the~possibilities for 
particle candidates remain quite wide: Models with very low or very high masses and/or very weak or strong interactions are still viable and escape current accelerator constraints. 
On the experimental side, the~sensitivity of neutrino telescopes, gamma- and X-ray telescopes and cosmic-ray detectors has improved enormously, sometimes by orders of magnitude, in~the last decades,
but the fact remains that no experiment has reported any positive signal so far. 
We seem to have past the point where a~clear, strong, unambiguous signal will be detected by any given experiment. It seems probable that, rather, any detection of particle dark
matter, if~that is the solution to our current observational conundrum, will show itself as a~slight deviation from the background. 
Given the uncertainties in the backgrounds of some of the techniques  mentioned in this review, it will not suffice that a~single experiment detects an~anomaly, but~that a~consistent
picture emerges from all the signatures that experimentalists are looking for. I~have mentioned Occam's razor a~couple of times in this review precisely because, taken individually, the~dark matter
explanation of any single observation might easily be superseded by an~hypothesis using only conventional physics. It is in the correlation between results of different experiments where
we should look. Paraphrasing Viktor Hess in his Nobel lecture in 1936: 
In order to make further progress, particularly in the field of dark  matter searches, it will be necessary to apply all our resources and apparatus simultaneously and side-by-side (while the original
formulation referred to the ``field of cosmic rays''~\cite{Hess:1936aaa}).  The~aim of this review is to give a~glimpse on the signatures looked for in indirect searches for dark matter,
accompanying them with examples of recent results and hopefully conveying to the reader that indirect searches for dark matter is a~complex and multidisciplinary field, currently~driven by
observations.

Searches for the products of dark matter annihilations in our galaxy or nearby galaxies with gamma-ray telescopes provide stringent limits to the dark matter annihilation cross
section, mainly~at relatively low dark matter masses, in~the region of a~few GeV to $\sim$100 GeV, but~they suffer from complex backgrounds. Dwarf spheroid galaxies are a~relatively
background free sources that are commonly used in these kind of searches. X-ray telescopes extend the mass reach of gamma-ray telescopes to the few keV region since they can detect softer
signatures from final-state radiation of charged decay or annihilation products of low mass candidates. Neutrino telescopes provide a~complementary search approach, currently more sensitive to high dark matter
masses, $\mathcal{O}$(100~GeV)  and above, while~searches with charged cosmic-rays suffer from uncertainties from propagation and relevant cross sections, like~antiproton
production cross section, needed to understand the production mechanisms at the source and as a~background. However, they also provide competitive results, comparable to the two
other techniques just mentioned. Large radio arrays have recently joined the effort of searching for annihilations of dark matter in dwarf galaxies by looking for anomalous
radio emission from synchrotron radiation of $e^+$ and $e^-$ produced as a~result of dark matter annihilations or decays. There are new astrophysical uncertainties involved in this kind of
technique due to the needed knowledge of the magnetic field of the source and the propagation of the electrons in the complex medium that can be a~galactic halo. However, results look promising,
and they are competitive with the results from gamma-rays, neutrinos and cosmic-rays under certain assumptions of the structure of the source. So these five, let us call them {\em source-based},
searches provide really complementary results and, taken together, will really help characterizing any~signal.

On top of the source-based searches, early universe cosmology presents an~independent way to search for dark matter, based on the effects of dark matter annihilations
(or decay) in the evolution of the early universe. Additional energy deposition on the primordial plasma, from~the CMB era to the first formation of structures, must not
cause a~distortion of the large-scale structure of the universe as we see it today. Quite stringent limits on the annihilation or decay characteristics of dark matter
can be extracted from such~requirement.

In the big picture of things the question remains: Is the dark sector really so simple as one stable particle species, while the visible sector comprises several
fundamental particles and families? Or,~even more radically: Is really the solution to the dark matter problem a~particle-based one? From a~pure experimental point of view,
there is currently no evidence for or against either of these options. The~particle solution was historically adopted since extensions of the Standard Model of particle physics
predict the existence of new particles, and~some of them have the characteristics needed of a~dark matter candidate (the ``WIMP miracle''), and~in order to reduce the 
parameter space, experimental searches are simply interpreted in terms of a~single particle solution. The~constraining power of the results presented in this review would
be lost if several candidates would have to be taken into account. Not least, the~hierarchy of the masses, strength of interactions and annihilation branching ratios of the
different candidates would be unknown. This poses a~limitation from the experimental point of view, but~it should not stop theorists from exploring such 
scenarios~\cite{Belanger:2020hyh,Yaguna:2019cvp,Elahi:2019jeo,Poulin:2018kap,Ahmed:2017dbb,Aoki:2012ub,Zurek:2008qg}. 

As for the second question above, there are proposed ways out from the dark matter conundrum without the need of dark matter.  
Modified Newtonian Dynamics (MOND)~\cite{Famaey:2011kh}, and~extensions thereof, is the more completely developed approach so far, but~it requires a~modification of
Newtonian gravity (or~General Relativity for that matter), theories that are understood to be universally valid and, therefore, any proposed modification is, at~a minimum,
philosophically difficult to swallow. MOND~really solves the problem it set out to solve: Explaining the flat rotation curves of galaxies taking into account only the
visible baryonic matter. It tends to fail, though, when trying to describe complex halos with velocity streams perpendicular to the galactic plane~\cite{Lisanti:2018qam},
and has serious difficulties with early universe cosmology, the~formation of large scale structure and, specifically, the~speed of gravitational waves~\cite{McGaugh:2014nsa,Sanders:2018jiv}.
Not that $\Lambda$CDM cosmology is free from problems
(a singularity at $t=0$ and the need for an~ad-hoc cosmological constant for example; not to mention the need for a~new type of dark matter itself).  It is not the aim to review here the
advantages of $\Lambda$CDM cosmology over alternative theories of gravity or vice-versa, but~I would just like to finish by reminding the reader that the lack of any
experimental signal so far from dark matter searches really forces us to keep all our options~open.

\vspace{6pt} 


\end{document}